\documentclass[fleqn,usenatbib]{mnras}

\usepackage{newtxtext,newtxmath}
\usepackage{booktabs}

\usepackage[utf8]{inputenc}
\usepackage[T1]{fontenc}
\usepackage{ae,aecompl}

\usepackage{graphicx}	
\usepackage{amsmath}	
\usepackage{amssymb}	
\usepackage[dvipsnames]{xcolor}
\usepackage{tabu}
\usepackage{longtable}

\newcommand{\Msun}{\ensuremath{M_{\odot}}}
 
\newcommand{\braket}[2]{\left< #1 \vphantom{#2} \right|
 \left. #2 \vphantom{#1} \right>} 
\let\baraccent=\= 
\renewcommand{\=}[1]{\stackrel{#1}{=}} 

\title[Emulating galaxy clustering and galaxy-galaxy lensing]{Emulating galaxy clustering and galaxy-galaxy lensing into the deeply
  nonlinear regime: methodology, information, and forecasts}

\author[B. D. Wibking et al.]{Benjamin D. Wibking$^{1}$\thanks{E-mail: wibking.1@osu.edu},
Andrés N. Salcedo$^{1}$,
David H. Weinberg$^{1}$,\newauthor
Lehman H. Garrison,$^{2}$,
Douglas Ferrer$^{2}$,
Jeremy Tinker$^{3}$,
Daniel Eisenstein$^{2}$,\newauthor
Marc Metchnik$^{4}$,
and Philip Pinto$^{4}$
\\
$^{1}$Dept. of Astronomy and Center for Cosmology and AstroParticle Physics, Ohio State University, 140 W 18th Ave, Columbus, OH, USA\\
$^{2}$Harvard-Smithsonian Center for Astrophysics, 60 Garden St., MS-10, Cambridge, MA 02138\\
$^{3}$Center for Cosmology and Particle Physics, New York University,
4 Washington Place, New York, NY 10003\\
$^{4}$Steward Observatory, University of Arizona, 933 N. Cherry Ave., Tucson, AZ 85121
}

\date{Accepted XXX. Received YYY; in original form ZZZ}

\pubyear{2017}

\begin{document}
\label{firstpage}
\pagerange{\pageref{firstpage}--\pageref{lastpage}}
\maketitle

\begin{abstract}
The combination of galaxy-galaxy lensing (GGL) with galaxy clustering is one
of the most promising routes to determining the amplitude of matter clustering
at low redshifts. We show that extending clustering+GGL analyses from
the linear regime down to $\sim 0.5 \, h^{-1}$ Mpc scales increases their constraining power
considerably, even after marginalizing over a flexible model of non-linear
galaxy bias. Using a grid of cosmological N-body simulations, we construct a
Taylor-expansion emulator that predicts the galaxy autocorrelation $\xi_{\text{gg}}(r)$
and galaxy-matter cross-correlation $\xi_{\text{gm}}(r)$ as a function of $\sigma_8$,
$\Omega_m$, and halo occupation distribution (HOD) parameters, which are allowed
to vary with large scale environment to represent possible effects of galaxy
assembly bias. We present forecasts for a fiducial case that corresponds to
BOSS LOWZ galaxy clustering and SDSS-depth weak lensing (effective source
density $\sim 0.3$ arcmin$^{-2}$). Using tangential shear and projected
correlation function measurements over $0.5 \leq r_p \leq 30 \,
h^{-1}$ Mpc yields a 1.8\% constraint on the parameter combination $\sigma_8\Omega_m^{0.58}$, a factor of
two better than a constraint that excludes non-linear scales ($r_p > 2
\, h^{-1}$ Mpc,
$4 \, h^{-1}$ Mpc for $\gamma_t,w_p$). Much of this improvement comes
from the non-linear clustering information, which breaks degeneracies
among HOD parameters that would otherwise degrade the inference of
matter clustering from GGL. Increasing the effective source density to $3$ arcmin$^{-2}$ sharpens the constraint on
$\sigma_8\Omega_m^{0.58}$ by a further factor of two. With robust
modeling into the non-linear regime, low-redshift measurements of matter clustering
at the 1-percent level with clustering+GGL alone  are well within
reach of current data sets such as those provided by the Dark Energy Survey. 
\end{abstract}




\section{Introduction}
Weak gravitational lensing is the most powerful tool for measuring the
clustering of dark matter at low redshifts. Cosmic shear analyses use
the correlated ellipticities of lensed galaxies to infer the power
spectrum of foreground mass fluctuations. In galaxy-galaxy lensing
one correlates a shear map with the distribution of foreground
galaxies to infer the galaxy-matter cross-correlation. This
cross-correlation probes the halo mass profiles and dark matter
environments of different classes of galaxies (e.g., \citealt{Mandelbaum_2006}), a valuable diagnostic of galaxy formation physics. The
cross-correlations can be combined with measurements of galaxy
clustering to infer the amplitude of matter clustering and thereby test
dark energy or modified gravity theories for the origin of cosmic
acceleration \citep{Weinberg_2013}.

\begin{figure*}
  \includegraphics[width=\textwidth]{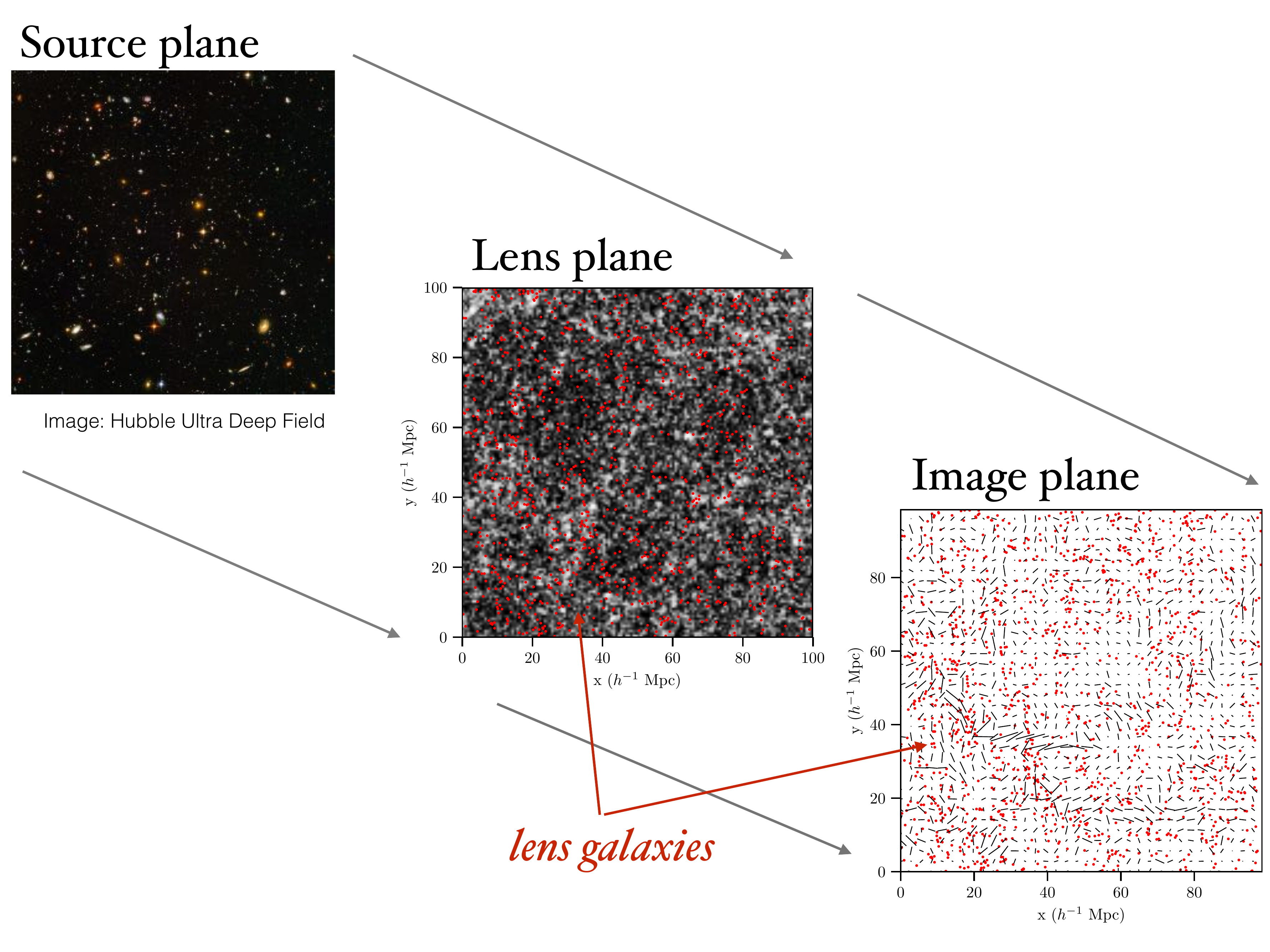}
  \caption{Pictorial illustration of clustering+GGL. The central panel
    shows the galaxy distribution (red points) and projected matter
    distribution (grey scale) in a $400 \, h^{-1}$ Mpc slice through our
    fiducial simulation at $z=0.3$. More distant source galaxies
    (illustrated here by the Hubble Ultra-Deep Field) are sheared by
    the intervening dark matter. The cross-correlation of the galaxy
    distribution and the shear field (shown by lines in the right
    panel, with arbitrary normalization) is the GGL signal, which can
    be combined with the galaxy clustering to infer the dark matter
    clustering at the redshift of the lens population.}
  \label{fig:cartoon}
\end{figure*}

The opportunity is easy to understand at the level of linear
perturbation theory, which should describe matter clustering and
galaxy bias on large scales where clustering is weak. In this regime,
the galaxy and matter auto-correlations are related by a
scale-independent bias factor, $\xi_{\text{gg}}(r) = b_g^2 \xi_{\text{mm}}(r)$. The
galaxy-galaxy lensing (hereafter GGL) signal is proportional to
$\Omega_m \xi_{\text{gm}} (r) = \Omega_m r_{\text{gm}} b_g \xi_{\text{mm}}(r)$, where $\Omega_m$ is
the matter density parameter and the galaxy-matter cross-correlation
coefficient $r_{\text{gm}}$ is expected to approach one on large
scales. Assuming $r_{\text{gm}} = 1$, one can combine the GGL and
$\xi_{\text{gg}}(r)$ measurements to cancel the unknown bias factor $b_g$ and
constrain $\Omega_m \sqrt{\xi_{\text{mm}}(r)}$. The amplitude of this
observable can be summarized by the product $\sigma_8 \Omega_m$, where $\sigma_8$ is the rms linear theory matter
fluctuation in $8 \, h^{-1}$ Mpc spheres. In practice, the best
constrained parameter combination differs from $\sigma_8 \Omega_m$
because the value of $\Omega_m$ affects the shape of the matter
correlation function and because geometric distance factors that enter
the lensing signal depend on $\Omega_m$ (see discussion in \citealt{Jain_1997} and in section
\ref{section:observables} below).  We illustrate the GGL measurement
pictorially in Figure \ref{fig:cartoon}.  In this paper, we use cosmological $N$-body simulations and halo occupation distribution (HOD;
\citealt{Berlind_2002}) methods to predict galaxy clustering and
GGL into the deeply nonlinear regime, where $b_g$ may become
scale-dependent and $r_{\text{gm}}$ may depart from unity. We illustrate the
considerable gains that can be made by exploiting small scale GGL and
$\xi_{\text{gg}}$ measurements in these analyses.

Several previous studies have investigated the use of HODs or related
methods to model GGL and galaxy clustering into the non-linear regime
(\citealt{Yoo_2006,Leauthaud_2011,Cacciato_2012,Yoo_2012,More_2013a,More_2013b})
and sharpen the resulting cosmological constraints. These studies have
generally relied on analytic approximations with some numerical
simulation tests, but the precision of observations has reached the
point that the accuracy of the analytic approximations is becoming a
limiting factor. Our approach is similar in spirit to the numerically
based ``emulator'' scheme introduced by \cite{Heitmann_2009} to predict
nonlinear matter power spectra, extended here with HOD parameters to
predict $\xi_{\text{gm}}$ and $\xi_{\text{gg}}$. We ultimately plan to consider a
grid of cosmological parameters that spans the space allowed by cosmic
microwave background (CMB) data, but in this paper we consider a
fiducial cosmology based on Planck CMB results \citep{Planck_2016} plus four
simulations with fixed steps in $\sigma_8$ and $\Omega_m$ (at fixed
$\Omega_m h^2$). For our fiducial HOD, we consider parameters
appropriate to the LOWZ sample of the Baryon Oscillation Spectroscopic
Survey (BOSS; \citealt{Eisenstein_2011}; \citealt{Dawson_2013}), as the
combination of imaging from the Sloan Digital Sky Survey (SDSS; \citealt{York_2000}) with BOSS LOWZ spectroscopy is one of the most powerful
current data sets for clustering and GGL analysis \citep{Singh_2016}. Instead of the Gaussian Process emulator of \cite{Heitmann_2009}, we use a simple linear Taylor expansion in cosmological and HOD
parameters. This approach becomes viable when the observational
constraints about fiducial parameters are tight, but its adequacy must
be tested in the context of any specified data analysis.

HOD methods characterize the relation between galaxies and dark matter
in terms of the probability $P(N|M_{\text{halo}})$ that a halo of mass $M_{\text{halo}}$ contains
$N$ galaxies of a specified class \citep{Benson_2000}. The principal
question for cosmological inference from GGL and clustering is whether
the adopted HOD parameterization has enough freedom to represent
non-linear galaxy bias at the level of accuracy required in order to model the
observations. The clustering of dark matter halos depends on their
formation history as well as their mass, an effect commonly known as
halo assembly bias (\citealt{Sheth_2004}; \citealt{Gao_2005}; \citealt{Harker_2006}; \citealt{Wechsler_2006}). Correlations of galaxy properties
with halo assembly history at fixed mass can therefore induce galaxy
assembly bias, which is not accounted for in traditional HOD
parameterizations. Analyzing the mock galaxy catalogs of \cite{Hearin_2013}, \cite{McEwen_2016} show that even when the
model galaxy population has substantial assembly bias, fitting it with
a standard HOD yields a cross-correlation coefficient $r_{\text{gm}}(r)$
accurate at the $\sim$2 per cent level and thus predicts the correct
relative amplitude of galaxy clustering and GGL. In this paper, we
explicitly allow for variation of the HOD with large scale environment
in our parameterization, as a way of accounting for galaxy assembly
bias (see section \ref{section:assembly_bias}).

GGL measurements can be made with the same imaging data sets acquired
for cosmic shear analyses, though there are advantages to combining
deep imaging data with a spectroscopic survey of galaxies that serve
as the lensing sample. \cite{Mandelbaum_2013} analyzed galaxy
clustering and GGL in the SDSS DR7 \citep{Abazajian_2009} data set, restricting their
analysis to large scales where one can expect $r_{\text{gm}} = 1$. They found
$\sigma_8 (\Omega_m/0.25)^{0.57} = 0.80 \pm 0.05$, which can be scaled
to the now commonly used parameter $S_8 = \sigma_8 (\Omega_m /
0.3)^{0.5} = 0.72 \pm 0.05$ (where we ignore the small difference
between 0.5 and 0.57 in the exponent). This is lower than the value
$S_8 = 0.83 \pm 0.012$ inferred for a $\Lambda$CDM model
normalized to the Planck 2015 CMB data (\citealt{Planck_2016}, Table
4, TT+TE+EE+lowP+lensing column).

Many but not all recent cosmic
shear analyses also find low amplitudes for matter clustering compared to
the Planck value
(e.g. \citealt{Heymans_2012}; \citealt{Hildebrandt_2017}; but see \citealt{Jee_2016}). \cite{More_2015} use an analytic HOD-based approach to
model clustering of the BOSS CMASS galaxy sample (effective redshift
$z = 0.57$) and GGL measurements of CMASS from the 105 deg$^2$ of
overlap between BOSS and the CFHTLens imaging survey \citep{Heymans_2012}. Their results are consistent with Planck-normalized $\Lambda$CDM
predictions, but the errors are fairly large because of the limited
overlap area. Most recently, the Dark Energy Survey (DES)
Collaboration has derived $S_8 = 0.783^{+0.021}_{-0.025}$ from the
combination of clustering, GGL, \emph{and} cosmic shear in their Year
1 data set, weakening but not eliminating the tension with Planck
$\Lambda$CDM predictions (\citealt{DESY1KP}, Table II).

Particularly relevant to this paper, \cite{Leauthaud_2017} find
discrepancies of 20-40 per cent, well above their statistical errors, at scales $r < 10 \, h^{-1}$ Mpc between
their measurements of GGL for CMASS galaxies (from CFHTLens and SDSS
Stripe 82 imaging) and the numerical predictions from
Planck-normalized mock catalogs that reproduce observed CMASS galaxy
clustering. The stakes for robust modeling of nonlinear galaxy
clustering and GGL are therefore high, and the prospects for
high-precision measurements over a range of redshifts will grow
rapidly with future DES analyses and forthcoming data from the
Subaru Hyper-Suprime Camera (HSC) \citep{HSC2017}.

The next section describes the construction of our emulator, including
the simulation suite, our HOD prescription and formulation of assembly
bias, and the sensitivity of clustering and GGL observables to
parameter variations about our fiducial choices. In section \ref{section:forecast}, we
derive forecasts for constraints on $\sigma_8$, $\Omega_m$, and HOD
parameters, using covariance matrices appropriate to BOSS LOWZ galaxy
clustering and SDSS-depth GGL measurements \citep{Singh_2016}. We
show how the expected constraints depend on the choice of scales in
the galaxy clustering and GGL measurements and on the effective source
density of the weak lensing map. In section \ref{section:discussion}, we discuss the
implications of our results and the prospects for applying our
methodology to current and near-future data sets.

\section{Emulator Construction}
\label{section:emulator}
Our goal is to provide numerically calibrated analytic recipes to
compute the real-space matter auto-correlation $\xi_{\text{mm}}(r)$, galaxy
auto-correlation $\xi_{\text{gg}}(r)$, and galaxy-matter cross-correlation
$\xi_{\text{gm}}(r)$ for cosmological parameters and HOD parameters that are
perturbations around a fiducial model. From these one can compute
projected observables that are directly measurable in a galaxy
redshift survey or weak lensing survey (see section \ref{section:observables} below). Our
fiducial cosmological parameters, based on the Planck 2015 CMB
analysis \citep{Planck_2016} are $\Omega_m = 0.3142$, $\Omega_b = 0.0491$, $h =
0.6726$, $n_s = 0.9652$, and a linear theory power spectrum normalization at
$z=0$ of $\sigma_8 = 0.83$. We assume a flat universe with a
cosmological constant and three massless neutrino species
($N_{\text{eff}}=3.04$) with zero cosmological neutrino density. In this paper we consider variations of $\sigma_8$
and $\Omega_m$, the two parameters that most affect the relative
amplitude of galaxy clustering and GGL. When varying $\Omega_m$ we
hold $\Omega_m h^2$, $\Omega_b h^2$, $n_s$, and $\sigma_8$ fixed.

For fiducial HOD parameters we choose values appropriate to the BOSS
LOWZ galaxy sample at median redshift $z=0.27$ (see section
\ref{section:hod}). Our methods can be readily
extended to other galaxy samples and other redshifts chosen on the
basis of specified observational data sets.

\subsection{Numerical simulations}
Our simulation procedures are described in detail by \cite{Garrison_2017}.
They use the \textsc{Abacus} $N$-body code (Ferrer et al., in prep.;
Metchnik \& Pinto, in prep.; see also \citealt{Metchnik_2009}) and
initial conditions computed with the configuration-space 2LPT code
described by \cite{Garrison_2016}. The input power spectra were
generated by the linear Boltzmann code \textsc{CAMB} \citep{Lewis_2011} for
redshift $z=0$ and rescaled by the linear growth factor to the
starting redshift $z=49$.

For the results in this paper, we use the simulations that consist of
a fiducial cosmology favored by the Planck 2015 results, two variations in $\Omega_m$ ($0.2879$, $0.3442$) at fixed $\Omega_m h^2$, and two variations in
$\sigma_8$ ($0.78$, $0.88$). All of these simulations use the same
phases in their initial conditions in order to minimize cosmic
variance in the computation of derivatives (see section
\ref{section:observables}). These simulations have a box size of $720$ $h^{-1}$
Mpc, a particle mass of $1.09 \times 10^{10}$ $h^{-1}$ \Msun (for the
fiducial $\Omega_m$), with
$1440^3$ particles, and a Plummer softening length of $41 \, h^{-1}$
kpc. We use the particle outputs at redshift $z=0.3$ in this work,
close to the central redshift of BOSS LOWZ \citep{Parejko_2013,Tojeiro_2014}. We
refer to distances and densities in comoving units throughout this paper.

\subsection{Halo identification}
We identify halos using the \textsc{Rockstar} halo finder
\citep{Behroozi_2013}. However, we use strict (i.e., without
unbinding) spherical overdensity (SO) halo
masses around the halo centers identified with \textsc{Rockstar}, rather than
the default 6D FOF-like masses output by \textsc{Rockstar}.  For finding halos
we use a primary mass definition set to the virial mass of
\cite{Bryan_1998}, but for all halo masses used after halo finding is
complete we adopt the $M_{200b}$ mass definition, i.e., the mass
enclosed by a spherical overdensity of
200 times the mean matter density at a given redshift and
cosmology. Thus, although \textsc{Rockstar} is our identification tool, our
eventual halo population consists of dark matter systems with
masses and radii defined by the $200 \rho_b$ criterion, effectively centered on
local peaks of the dark matter density. We do not make use of dark matter subhalos contained within larger halos, although subhalo masses are always included in parent halo masses.

We have found that the reported concentration parameters ($c =
R_{\text{halo}}/R_s$ for an NFW profile;  \citealt{NFW_1997}) are not reliable at the
mass and force resolution available in our simulations. To obtain
concentrations for creating satellite galaxy distributions, therefore,
we use the fitting formula of \cite{Correa_2015}, calibrated to
significantly higher resolution simulations. We perform an approximate rescaling from the halo masses defined by 200 times the critical density ($M_{200c}$) used there to the $M_{200b}$ definition as employed in this work.

\subsection{HOD prescription}
\label{section:hod}
Our HOD model is similar to that introduced by \cite{Zheng_2005} and
used in many galaxy clustering analyses (e.g., \citealt{Zehavi_2005},
\citealt{Zehavi_2011}; \citealt{Coupon_2012}; \citealt{Zu_2015}). Each
halo of a given mass can host central galaxies and satellite galaxies,
where hosting satellite galaxies is conditioned on hosting a central
galaxy. We compute the expectation value that a given halo will host a central
galaxy according to:
\begin{equation}
\braket{N_{\text{cen}}}{M_{\text{halo}}} = \frac{1}{2} \left[ 1 + \text{erf} \left(
\frac{\log M_{\text{halo}} - \log M_{\text{min}}}{\sigma_{\log M}} \right) \right],
\label{eq:hodcens}
\end{equation}
where $M_{\text{min}}$ is the
halo mass for which the occupation probability is one-half, and
$\sigma_{\log M}$ allows for logarithmic scatter between galaxy
luminosity and halo mass.  Throughout this paper we use $\log$ for
the base-10 logarithm and $\ln$ to indicate the natural logarithm.  The
number of central galaxies is always zero or one, chosen
stochastically given the expectation value. The central galaxy is
placed at the \textsc{Rockstar}-identified halo center. We compute the expectation value of the number of satellite
galaxies according to:
\begin{equation}
\braket{N_{\text{sat}}}{M_{\text{halo}}, N_{\text{cen}} = 1} = 
  \begin{cases}
    \left(\frac{M_{\text{halo}} - M_0}{M_1}\right)^{\alpha} & \text{ if } M_{\text{halo}} > M_0 \\
    0 & \text{ otherwise }.
  \end{cases}
\label{eq:hodsats}
\end{equation}
Here $M_0$ is a halo mass below which there are no satellite
galaxies, $M_1 + M_0$ is the halo mass for which there is
an average of one satellite galaxy, and $\alpha$ is the power-law slope
of the number of satellites as a function of halo mass. Satellite
galaxy counts in a given halo are sampled from a Poisson distribution with this expectation
value to populate individual halos. The positions of satellite
galaxies are chosen by sampling from a satellite galaxy profile:
\begin{equation}
\rho_{\text{sat}}(r) = r^{\Delta \gamma} \rho_{\text{NFW}}(r) \, ,
\end{equation}
where $\rho_{\text{NFW}}(r)$ is the NFW density profile with
concentration computed from the fitting formula of
\cite{Correa_2015} and the parameter $\Delta \gamma$ allows a
power-law deviation between the satellite galaxy profile and the NFW
profile of the mass distribution. We truncate the satellite density profile at the $R_{200,b}$ radius.

In an HOD analysis, the number density of galaxies is an important
constraint in addition to the galaxy clustering. For our emulator and
forecasts, we have elected to take $n_{\text{gal}}$ as an HOD parameter in
place of $M_{\text{min}}$. Once other parameters have been specified, we use
equations \ref{eq:hodcens} and \ref{eq:hodsats} to find the value of $M_{\text{min}}$ that yields the
specified $n_{\text{gal}}$, keeping the ratios $M_0 / M_{\text{min}}$ and $M_1 /
M_{\text{min}}$ fixed. The value of $M_0 / M_{\text{min}}$ is often ill-constrained
in HOD fits because it has negligible impact on number density or
clustering for $M_0 / M_1 \ll 1$. In this paper we have
chosen to fix $M_0 / M_1 = 0.089$ and not treat it as a free parameter;
our results would be negligibly different if we set $M_0 = 0$. Our set
of adjustable HOD parameters is therefore $n_{\text{gal}}$, $\sigma_{\log M}$, $M_1 / M_{\text{min}}$,
$\alpha$, and $\Delta \gamma$. For our fiducial model we adopt the
values $n_{\text{gal}} = 3 \times 10^{-4} \, h^3 \,$Mpc$^{-3}$,  
$\sigma_{\log M} = 0.68$, $M_1/M_{\text{min}}=9.55$, $\alpha = 1.15$,
and $\Delta \gamma = 0$ with the number density based on the LOWZ
results of \cite{Parejko_2013} and the other parameter values based on the $M_r < -21$ results of
\cite{Zehavi_2011}. We choose the non-number density parameter values
from another sample because they are significantly
better constrained for the $M_r < -21$ sample but still consistent with the values inferred
by modeling by \cite{Parejko_2013} of the LOWZ sample itself.

\subsection{Modeling galaxy assembly bias}
\label{section:assembly_bias}
Part of the motivation for HOD descriptions of galaxy bias (see
e.g. \citealt{Berlind_2002}) was the expectation from the simplest
formulations of excursion set theory \citep{Bond_1991} that halo clustering should be independent of halo formation
history at fixed halo mass \citep{White_1994}. While
this prediction proved a good match to early $N$-body results \citep{Lemson_1999}, more detailed measurements with larger
simulations have revealed a variety of correlations between formation
history and halo clustering (e.g., \citealt{Sheth_2004,Gao_2005,Harker_2006,Wechsler_2006,Salcedo_2017}). These correlations can cause the galaxy HOD to vary
systematically with halo environment, in which case a calculation that
assumes a single global HOD will make incorrect predictions for galaxy
clustering and GGL. For example, a model in which galaxy stellar mass
is tied to halo peak circular velocity (rather than halo mass) and
galaxy color is tied to halo formation time exhibits significant
``galaxy assembly bias'' for samples defined by luminosity and color
cuts; correlation functions change significantly if galaxies
are shuffled among halos of the same mass in a way that erases
correlations with halo assembly \citep{Zentner_2014}.

To allow for assembly bias effects in our HOD model, we have
introduced a parameter $Q_{\text{env}}$ that shifts the cutoff $M_{\text{min}}$ of
the central galaxy occupation as a function of the halo's large scale
environment. Specifically, we compute the overdensity $\delta_8$
around each halo in a top-hat sphere of radius $8 \, h^{-1}$ Mpc and rank
all halos (from 0 to 1) in order of increasing $\delta_8$ in narrow
(0.1 dex) bins of halo mass. We then choose an environment-dependent
$M_{\text{min}}$ for each halo according to
\begin{equation}
\log M_{\text{min}} = \log M_{\text{min},0} + Q_{\text{env}} \left[\text{rank}(\delta_8) - 0.5\right],
\end{equation}
with a halo at the median overdensity for its mass having $M_{\text{min}} =
M_{\text{min},0}$. This prescription is similar to that introduced by \cite{McEwen_2016}, but using halo rank instead of $\delta_8$
directly makes the result for a given $Q_{\text{env}}$ less dependent on the
specific choice of environmental variable. It is also fairly
intuitive, e.g., for $Q_{\text{env}}=0.1$ the halos at the environmental
extremes have $M_{\text{min}}$ across a range of 0.1 dex about that of
halos in the median environment. Because we fix $M_1 / M_{\text{min}}$, the satellite
occupation shifts in $\log M_{\text{halo}}$ together with the central occupation.

\subsection{Emulated quantities}
\label{section:observables}
We use \textsc{Corrfunc} \citep{Sinha_2017} to compute the real-space galaxy 
autocorrelation $\xi_{\text{gg}}$, galaxy-matter cross-correlation 
$\xi_\text{{gm}}$, and matter autocorrelation $\xi_{\text{mm}}$ on
scales $0.01 < r < 125 \, h^{-1}$ Mpc, averaging over 20 realizations of
the HOD at each point in parameter space. 
Separately, we compute the linear matter autocorrelation 
$\xi_{\text{mm, lin}}$ by computing 
the appropriate integral over the linear power spectrum used for the
initial conditions.

Our emulator uses finite differences to compute a linear Taylor
expansion for ratios of these quantities as a function of scale:
\begin{align}
&b_{\text{nl}} = \left[\frac{\xi_{\text{mm}}}{\xi_{\text{mm,lin}}}\right]^{1/2} \, , \\
&b_{\text{g}} =  \left[\frac{\xi_{\text{gg}}}{\xi_{\text{mm}}}\right]^{1/2} \, , \\
&r_{\text{gm}} = \left[\frac{\xi_{\text{gm}}^2}{\xi_{\text{gg}} \, \xi_{\text{mm}}}\right]^{1/2} \, .
\end{align}
If needed, we regularize the behavior of these functions on large
scales (where the correlation function measurements become noisy) so
they obey the expected limits $b_{\text{nl}}(r) \rightarrow 1$, $b_g(r)
\rightarrow \text{const.}$, and $r_{\text{gm}}(r) \rightarrow 1$ as $r \rightarrow \infty$.

We focus on ratios so that the influence of cosmological parameters is 
treated exactly in the linear regime; there is no need to use a 
numerical emulator to approximate the impact of parameter changes on 
the linear matter power spectrum (similar to the
methodology of \citealt{Mandelbaum_2013}, who used a linear Taylor
expansion in $b_{\text{nl}}^2$ in order to model the nonlinear matter correlation function). We expect this approach to give our 
emulator a wide range of validity, as the scale-dependence of 
non-linear corrections, galaxy bias, and $r_{\text{gm}}(r)$ should have a 
relatively weak dependence on parameters such as $\Omega_b$, $h$, and 
$n_s$. We will test this expectation using our larger simulation grid 
in future work. 

Our emulation formula is simply: 
\begin{equation}
X(r) = X_{\text{fid}}(r) + \sum\limits_{i} \Delta p_i \frac{\partial 
  X(r)}{\partial p_i}
\label{eq:emulation}
\end{equation}
where $X(r)$ may be $\ln b_{\text{nl}}(r)$, $\ln b_g(r)$, or $\ln r_{\text{gm}}(r)$, 
$X_{\text{fid}}(r)$ is the value in the fiducial model, $\Delta p_i =
p_i - p_{i,\text{fid}}$ is the difference in parameter $i$ between the 
emulated model and the fiducial model, and the derivatives are 
evaluated about the fiducial model. The specific parameters that we 
use are: $\ln \sigma_8$, $\ln \Omega_m$, $\ln n_{\text{gal}}$, $\ln 
\sigma_{\log M}$, $\ln M_1/M_{\text{min}}$, $\ln \alpha$, $\Delta 
\gamma$, and $Q_{\text{env}}$. We generally expect logarithmic derivatives to give a greater range of 
validity because they can represent power-law relations not just 
linear relations, but we use linear derivatives for $\Delta \gamma$
and $Q_{\text{env}}$ because their fiducial values are zero.

We compute the partial derivatives in equation \ref{eq:emulation} by centered finite 
differences with step sizes determined by our set of grid points in cosmological and HOD parameter space. The HOD parameter space used consists of individual parameter variations about the fiducial HOD (evaluated at the fiducial cosmology) at $n_{\text{gal}} = \{0.00027, 0.00033\}$, $\sigma_{\log M} = \{0.58, 0.78\}$, $M_1/M_{\text{min}} = \{9.05, 10.05\}$, $\alpha = \{1.0, 1.3\}$, $\Delta \gamma = \{-0.1, 0.1\}$ and $Q_{\text{env}} = \{-0.1, 0.1\}$. In Appendix \ref{appendix:derivatives}, we tabulate our values of $\xi_{\text{mm,lin}}(r)$, 
$b_{\text{nl,fid}}$, $b_{\text{g,fid}}$, $r_{\text{gm,fid}}$, and the 
partial derivatives, allowing anyone to reproduce our emulator predictions.

The direct observables that we wish to emulate are the projected 
galaxy correlation function $w_p(r_p)$ and the excess surface density 
$\Delta \Sigma(r_p)$. Neglecting sky curvature, residual 
redshift-space distortion, and higher-order lensing corrections, these 
are related to the 3D real-space correlation functions by the 
projection integrals 
\begin{align}
&w_p(r_p) = 2 \int_{0}^{\pi_{\text{max}}}
\xi_{\text{gg}}\left(\sqrt{r^2 + \pi^2}\right) \, d\pi \, , \label{eq:wp} \\
&\begin{aligned}
  \Delta \Sigma(r_p) = \bar\rho \, \left[ \frac{4}{r_p^2} \int_{0}^{r_p} r \int_{0}^{\infty} \xi_{\text{gm}}\left(\sqrt{r^2 + \pi^2}\right) \, d\pi dr \, \right. \\
  \left. - \, 2 \int_{0}^{\infty} \xi_{\text{gm}}\left(\sqrt{r_p^2 +
        \pi^2}\right) \, d\pi \right] \, , 
\end{aligned} \label{eq:ds}
\end{align}
where the cosmic mean matter density is given in comoving coordinates
\begin{equation}
\bar \rho = \Omega_m \left( \frac{3 H_0^2}{8 \pi G} \right).
\end{equation}
We report $w_p$ and $\Delta\Sigma$ in units of $h^{-1}$ and $h \, \Msun$ pc$^{-2}$, respectively.

We compute the 3D correlation functions from our emulator via 
\begin{align}
&\xi_{\text{gg}} = b_{\text{g}}^2 \left(b_{\text{nl}}^2 \xi_{\text{mm,
  lin}}\right) \, , \\
&\xi_{\text{gm}} = r_{\text{gm}} b_{\text{g}} \left(b_{\text{nl}}^2
  \xi_{\text{mm, lin}}\right) \, ,
\end{align}
where all quantities in both equations depend on the 3D separation
$r$.

The choice of the $\pi_{\text{max}}$ cutoff for computing $w_p(r_p)$
depends on the redshift survey analysis; ideally one would like 
$\pi_{\text{max}} \rightarrow \infty$ to eliminate redshift-space 
distortions entirely, but estimates of $w_p(r_p)$ can become noisy for 
very large $\pi_{\text{max}}$. In this paper, we choose $\pi_{\text{max}} = 100 \, h^{-1}$ Mpc.  We assume that the impact of 
residual redshift-space distortion is accounted for in the 
redshift survey analysis.  On the largest scales we consider, the
redshift-space correction to $w_p$ may be as large as 15 per cent for our chosen value of $\pi_{\text{max}}$ \citep{vdBosch_2013}.

The computation of $\Delta \Sigma$ from GGL 
observations depends on photometric redshift estimates for the source 
galaxies and on cosmological parameters used to compute lensing 
critical surface densities (e.g. \citealt{Mandelbaum_2005}). In a cosmological 
analysis one might instead use our emulator to predict the more directly 
observed mean tangential shear
\begin{equation}
\begin{split}
\gamma_{t}(\theta) = \int dz_{\text{lens}} \int dz_{\text{src}} \,
n_{\text{lens}}(z_{\text{lens}}) n_{\text{src}}(z_{\text{src}})
\Theta(z_{\text{src}}-z_{\text{lens}}) \\
\times \, \frac{\Delta\Sigma(\theta,
  z_{\text{lens}})}{\Sigma_c(z_{\text{lens}},z_{\text{src}})} \, ,
\end{split}
\label{eq:gamma_t}
\end{equation}
where the step function $\Theta(x)$ ensures that lensing contributions
occur only when $z_{\text{lens}} < z_{\text{src}}$. Equation
\ref{eq:gamma_t} can incorporate the cosmological dependence of
distance ratios, nuisance parameters for 
photometric redshift uncertainties, and any signal-to-noise weighting
applied to the observations (through additional factors modifying $\Delta\Sigma$). The (comoving) critical surface density $\Sigma_c$
is
\begin{equation}
\Sigma_c = \frac{c^2}{4 \pi G} \frac{D_C(z_{\text{src}})}{D_C(z_{\text{lens}})
\left[ D_C(z_{\text{src}}) - D_C(z_{\text{lens}}) \right] (1+z_{\text{lens}})}
\, ,
\end{equation}
where $D_C(z)$ denotes the comoving distance to redshift $z$.

For the purpose of the forecasts in this paper, we compute
$\gamma_t(r_p)$ predictions from our emulator assuming that
$n_{\text{lens}}$ is a delta function centered at the effective lens
redshift $z_{\text{lens}} = 0.27$ (computed for BOSS LOWZ by
\citealt{Singh_2016}) and $n_{\text{src}}$ is a $\delta$-function
centered at an effective source redshift $z_{\text{src}} = 0.447$,
chosen so that the resulting critical lensing surface density is equal to the value $\Sigma_c = 4.7 \times 10^{3} \, h \,
\Msun$ pc$^{-2}$ given by \cite{Singh_2016}. Using $\gamma_t$ instead
of $\Delta\Sigma$ as the observable in our forecasts introduces an
additional dependence on cosmology that significantly modifies the
$\Omega_m$-$\sigma_8$ degeneracy direction, as the distances entering
$\Sigma_c$ involve an integral over the factor $[\Omega_m(1+z)^3 +
\Omega_{\Lambda}]^{-1/2}$. We find $d \ln
\Sigma_c(z_{\text{lens}},z_{\text{src}};\Omega_m) / d \ln \Omega_m
\approx 0.12$ at our fiducial values of $\Omega_m$, $z_{\text{lens}}$,
and $z_{\text{src}}$. Since $\gamma_t \propto \Delta \Sigma \,
\Sigma_c^{-1}$, this dependence modifies the best-constrained
cosmological parameter by a factor $\approx \Omega_m^{-0.12}$.
Because the amplitude of the lensing signal $\gamma_t$ has an additional dependence on $\Omega_m$ beyond that of $\Delta\Sigma$, using $\gamma_t$ marginally improves the constraining power of the measurement compared to assuming that $\Delta \Sigma$ is the observable.

Additionally, we correct for the $\Omega_m$-dependence of the projected distance as a function of angular separation and redshift $r_p(\theta, z; \Omega_m)$ by assuming the observer has estimated projected distances in an $\Omega_m = 0.3$ universe. We thus rescale the `true' distances in which we measure our correlation functions to those our observer would compute when calculating $w_p$ and $\Delta\Sigma$ (eqs. \ref{eq:wp} and \ref{eq:ds}). However, we find that this correction is very small and makes almost no difference to our results.

\section{Forecasting Constraints}
\label{section:forecast}
\subsection{Covariance matrices}
We use the following expressions from \cite{Singh_2016} for the
Gaussian component of the observable covariances: 
\begin{align}
&\begin{split}
\text{cov}_{w_p}(r_i,r_j) = \frac{2 A_{ij}}{A_i A_j} \int_{0}^{\infty}
\frac{k \, dk}{2 \pi} J_0(k r_i) J_0(k r_j) \left( b_g^2 P(k) +
  \frac{1}{n_{\text{g}}} \right)^2 \, ,
\end{split} \label{eq:wp_cov} \\
&\begin{split}
\text{cov}_{\Delta \Sigma}(r_i,r_j) = \frac{V_{ij}}{V_i V_j}
\int_{0}^{\infty} \frac{k \, dk}{2 \pi} J_2(k r_i) J_2(k r_j) \\
\times \, \left[ \left( b_g^2 P(k) + \frac{1}{n_{\text{g}}} \right) \left(
    \Delta \Pi \bar{\rho}^2 P(k) + \frac{\Sigma_c^2 \sigma_{\gamma}^2
    }{ n_{\text{s}}} \right) + \Delta \Pi \left(\bar{\rho}
    b_{\text{g}} r_{\text{gm}} P(k)\right)^2 \right] \, .
\end{split} \label{eq:DS_cov}
\end{align}
These expressions neglect line-of-sight modes, 
terms that can arise from using a suboptimal estimator that does not
subtract the tangential shear around random points, and redshift
evolution over the lensed galaxy population. We convert the $\Delta
\Sigma$ covariance to our lensing observable $\gamma_t$ covariance by
\begin{equation}
\text{cov}_{\gamma_t}(r_i,r_j) = \text{cov}_{\Delta \Sigma}(r_i,r_j) \, \Sigma_c^{-2}.
\end{equation}
The area and volume normalization factors in these expressions are: 
\begin{align}
A_{ij} &= \int_{0}^{\infty} \frac{k \, dk}{2 \pi} J_0(k r_i) J_0(k r_j)
  [W(k)]^2 \, ,\\
A_i &= \int_{0}^{\infty} \frac{k \, dk}{2 \pi} J_0(k r_i) [W(k)]^2 \, , \\
V_{ij} &= L_W A_{ij} \, ,\\
V_i &= L_W A_i \, .
\end{align}
Following \cite{Singh_2016}, we adopt survey
parameters appropriate to BOSS LOWZ galaxy-galaxy lensing.
The window function corresponds to a circular survey on the sky 
of radius $R_s = 1275 \, h^{-1}$ Mpc,
\begin{equation}
W(k) = 2 \pi R_{s}^2 \frac{J_1(k R_{s})}{k R_{s}},
\end{equation}
where $n_g = 3 \times 10^{-4} \, h^{3}$ Mpc$^{-3}$ is the galaxy number density, $\Delta \Pi = 400 \, h^{-1}$ Mpc is the 
effective line-of-sight lensing distance, $L_W = 500 \, h^{-1}$ Mpc is the effective 
line-of-sight survey window, $\Sigma_c = 4.7 \times 10^{3} \, h \, \Msun$ pc$^{-2}$ is the critical 
lensing surface density, $\sigma_{\gamma} = 0.21$ is the shape noise per galaxy, $n_s = 8 \, h^2$ Mpc$^{-2}$
is the effective projected number density of source galaxies, $\bar
\rho = \Omega_m (3H_0^2 / 8\pi G)$ is the cosmic mean matter density
(in comoving coordinates), and $P(k)$ is the nonlinear 
matter power spectrum as computed from our fiducial simulation. At our
adopted lens redshift, the survey area is 9000 deg$^2$ and the nominal
source
density is 1 arcmin$^{-2}$. This nominal value is reduced to an effective source
density of 0.3 arcmin$^{-2}$ due to redshift cuts (contributing
a $\sim 50$
per cent reduction) and by signal-to-noise
weighting (contributing a $\sim 30$ per cent reduction) (S. Singh, personal communication).

\begin{figure*}
  \includegraphics[width=\textwidth]{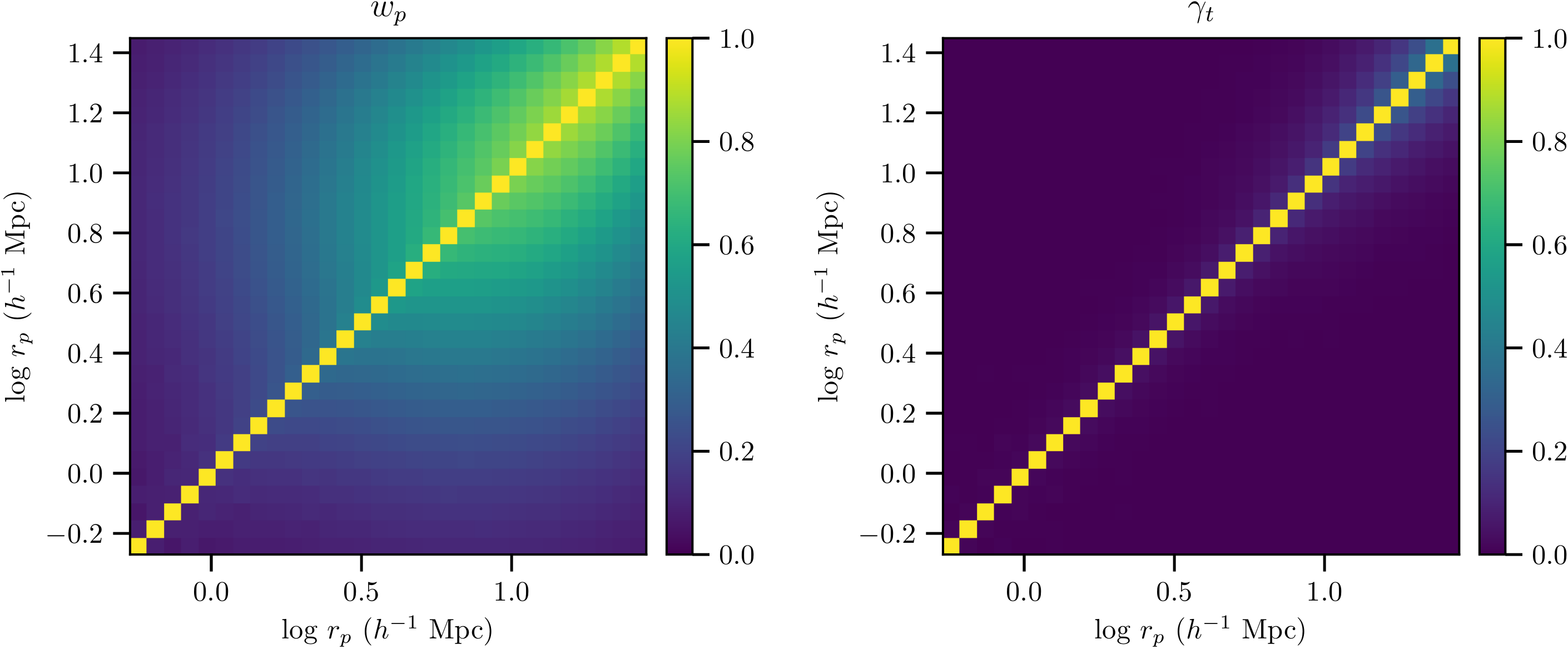}
  \caption{The correlation matrices for our forecast. We compute the covariance matrix for $\gamma_t$ with the integrals for the Gaussian contributions to the covariance for the 
 variance-minimizing estimator described by Singh et al. (2016). We use 
 similar integrals to compute the covariance for clustering. We assume 
 a source density, area, and redshift properties similar to those of the BOSS LOWZ 
 spectroscopic sample for clustering measurements and SDSS imaging 
 for lensing source galaxies.}
\label{fig:cov}
\end{figure*}

The correlation matrices 
\begin{equation}
\text{corr}(r_i, r_j) =
\frac{\text{cov}(r_i,r_j)}{\sqrt{\text{cov}(r_i,r_i) \, \text{cov}(r_j,r_j)}}
\end{equation}
for our fiducial forecast are shown in Figure \ref{fig:cov}.  The
$\gamma_t$ correlation matrix is nearly diagonal because of the
dominant contribution from shot noise, while the $w_p$ correlation
matrix has substantial off-diagonal terms for $r_p > 2 \, h^{-1}$
Mpc. For purposes of the forecast, we assume that non-Gaussian contributions to the covariance (e.g.,
\citealt{Scoccimarro_1999}; \citealt{Cooray_2001}) will be
minimized by masking the largest several clusters from the survey in
the clustering and GGL measurements used for cosmological analysis. We
also assume that the cross-observable covariance (i.e., the covariance between $\gamma_t$
and $w_p$) is negligible and the contribution to the covariance from uncertainties
in our knowledge of the true cosmic mean observables due to the finite
size of our simulations is negligible, although we plan to quantify
these contributions to the covariance in future
work.

\subsection{Model predictions}
\begin{figure*}
  \includegraphics[width=\textwidth]{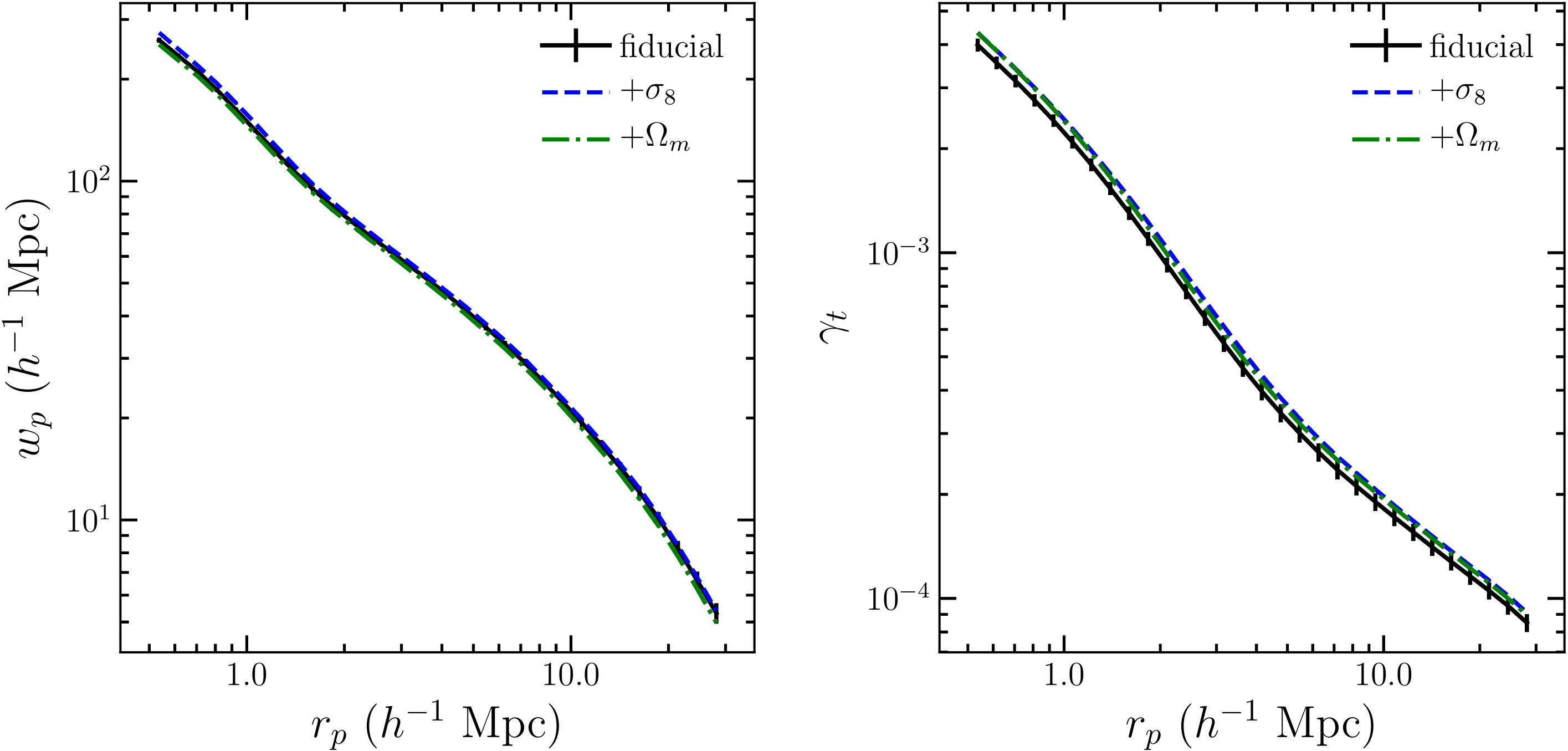}
  \caption{Projected correlation function (left) and tangential shear
    (right) predicted for our fiducial HOD parameters and the fiducial
    cosmological simulation (black solid) and for simulations with
    $\sigma_8$ increased by 6 per cent (blue dashed) or $\Omega_m$
    increased by 10 per cent (green dot-dashed). Error bars on the
    black curve show the diagonal errors for our assumed data
    properties, corresponding to BOSS LOWZ lens galaxies and
    SDSS-depth imaging. The magnitude of changes can be seen more
    clearly in Figures \ref{fig:derivatives} and \ref{fig:ratios}.}
  \label{fig:fiducial_emulator}
\end{figure*}
Figure \ref{fig:fiducial_emulator} shows the predicted $w_p(r)$ and $\gamma_t(r)$ from applying
our fiducial HOD to our fiducial cosmological simulation (black curve)
and to the simulations with higher $\sigma_8$ (blue dashed) and higher
$\Omega_m$ (green dot-dashed). On the scale of this figure, the impact of
these parameter changes (6 per cent in $\sigma_8$ and 10 per cent in
$\Omega_m$) is barely discernible, but one can see that the fractional
changes to $\gamma_t(r)$ are larger than the fractional changes to
$w_p(r)$. Increasing $\sigma_8$ boosts $\xi_{\text{mm}}(r)$, but at
fixed number density $n_{\text{gal}}$ the galaxy bias $b_g$
decreases. For $w_p \propto b_g^2 \xi_{\text{mm}}$ the effects nearly
cancel, while for $\gamma_t \propto b_g \xi_{\text{mm}}$ there is a
net increase of amplitude. Increasing $\Omega_m$ changes the shape of
$\xi_{\text{mm}}(r)$ and thus of $w_p(r)$, but the effect of a 10 per
cent change is subtle. Increasing $\Omega_m$ boosts the amplitude of
$\gamma_t(r)$ mainly by increasing the $\bar \rho$ pre-factor of
$\Delta\Sigma$ in equation \ref{eq:ds}; the change to $\Sigma_c$ in equation
\ref{eq:gamma_t} goes in the opposite direction but with much smaller
amplitude. The changes of observables are large compared to the
statistical errors expected for our fiducial data assumptions, but we
have not yet considered degeneracy between cosmological and HOD
parameters.

\begin{figure*}
  \includegraphics[width=\textwidth]{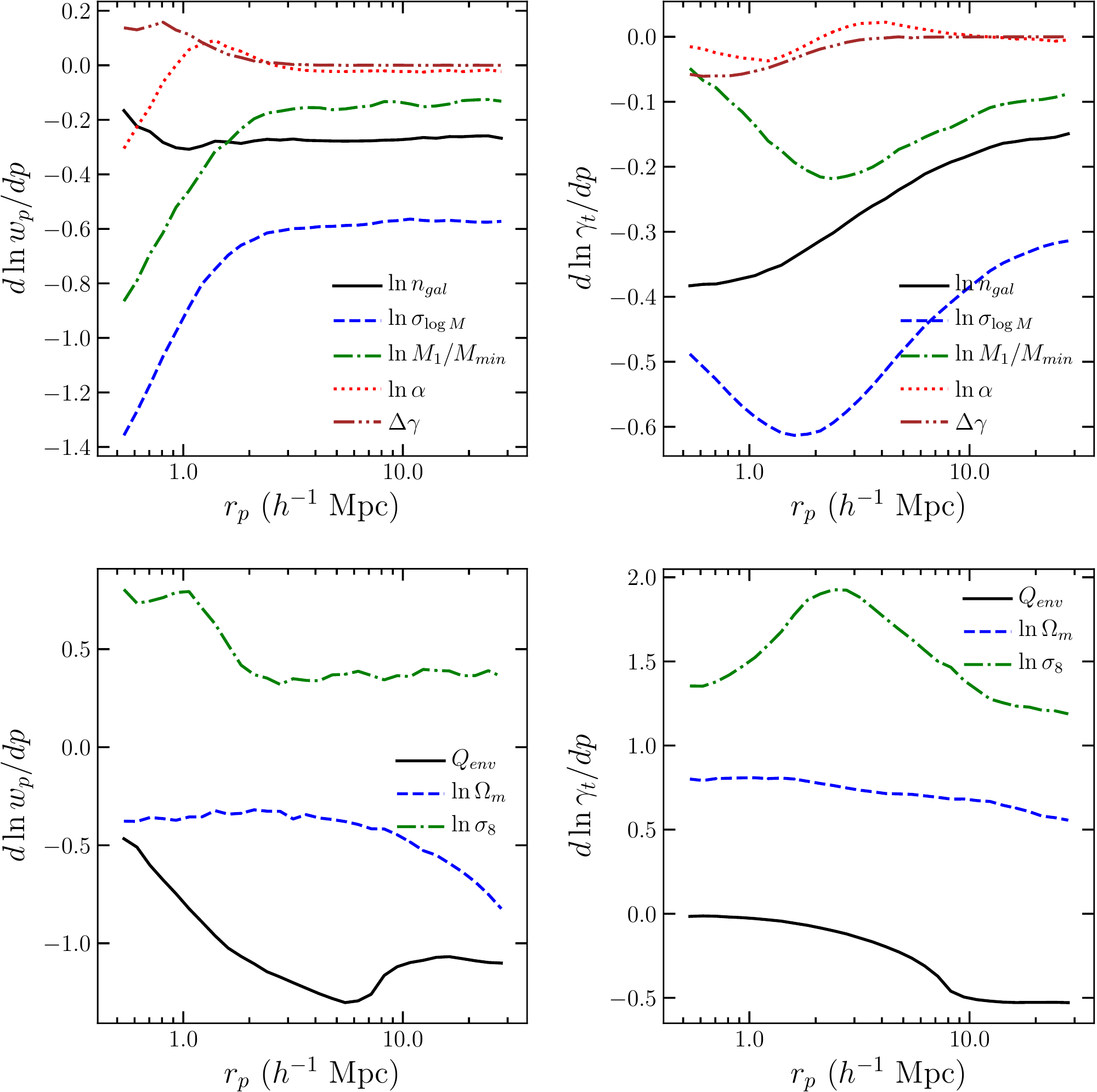}
  \caption{Logarithmic derivatives of $w_p$ (left panels) and
    $\gamma_t$ (right panels) with respect to HOD parameters and
    cosmological parameters, as indicated in the legends. Top panels
    show derivatives for standard HOD parameters. Bottom panels show
    derivatives for $\sigma_8$, $\Omega_m$, and the environmental HOD
    parameter $Q_{\text{env}}$.}
  \label{fig:derivatives}
\end{figure*}

Figure \ref{fig:derivatives} shows derivatives of $\ln w_p(r)$ (left panel) and $\ln
\gamma_t(r)$ (right panel) with respect to our eight model
parameters. The top panels show derivatives for the conventional HOD
parameters: $\ln n_{\text{gal}}$, $\ln \sigma_{\log M}$, $\ln
M_1/M_{\text{min}}$, $\ln \alpha$, and $\Delta \gamma$. For $w_p(r)$,
the large scale behavior is constant in $r_p$, corresponding to
changes in the asymptotic value of $b_g$, but the derivatives change
below $r_p \approx 2 \, h^{-1}$ Mpc as the 1-halo contributions to
$\xi_{\text{gg}}(r)$ become important. Thus, the parameters have
degenerate effects on linear scales, but using the full range of
$w_p(r)$ can break these degeneracies. Increasing $n_{\text{gal}}$
decreases $w_p(r_p)$ on all scales by shifting central galaxies to
less massive, more numerous, less biased halos. Increasing
$\sigma_{\log M}$ has a similar effect at large scales, and it
suppresses $w_p(r_p)$ more severely in the 1-halo regime because more
central galaxies reside in halos that are not massive enough to host
satellites. Increasing $M_1/M_{\text{min}}$ decreases the overall
fraction of satellites, depressing the large scale bias slightly and
the 1-halo correlations more severely. Increasing $\alpha$ at fixed
$M_1/M_{\text{min}}$ has almost no impact at large scales, but it
slightly boosts $w_p(r_p)$ on scales corresponding to the virial radii
of cluster mass halos, and it depresses $w_p(r_p)$ on small scales
where lower mass halos dominate the 1-halo regime. Increasing $\Delta
\gamma$ has no impact in the 2-halo regime, and it slightly boosts
$w_p(r_p)$ inside $1 \, h^{-1}$ Mpc by steepening satellite galaxy
profiles.

The influence of these parameters on $\gamma_t(r_p)$ is qualitatively
similar, but the scale dependence is more complex because $\Delta
\Sigma$ is an \emph{excess} surface density, $\Delta \Sigma (r_p) =
\bar \Sigma(< r_p) - \bar \Sigma (r_p)$ where $\bar \Sigma(< r_p)$ is
averaged over all radii smaller than $r_p$ (see, e.g. \citealt{Sheldon_2004}). Even at $r_p = 30 \, h^{-1}$ Mpc, the impact of
$n_{\text{gal}}$, $\sigma_{\log M}$, and $M_1/M_{\text{min}}$ has not reached
the scale-independence expected asymptotically at large $r_p$.

The lower panels of Figure \ref{fig:derivatives} show derivatives with respect to the
cosmological parameters $\ln \sigma_8$ and $\ln \Omega_m$ and our
environment-dependent HOD parameter $Q_{\text{env}}$. Increasing
$Q_{\text{env}}$ reduces the large scale galaxy bias with other HOD
parameters held fixed, because it increases $M_{\text{min}}$ (and thus
decreases galaxy numbers) for halos of a given mass in denser
environments. The effect of $Q_{\text{env}}$ becomes mildly
scale-dependent inside the radius $r = 8 \, h^{-1}$ Mpc that we are
using to define halo environment, and it decreases towards small
scales because the 1-halo regime of $\xi_{\text{gg}}(r)$ or
$\xi_{\text{gm}}(r)$ depends only on integrals over the halo mass
function and galaxy density profile (see \citealt{Berlind_2002},
eq. 11). For our purposes, the most important effect of
$Q_{\text{env}}$ is that it decouples the large scale bias from the
conventional HOD parameters, so one cannot simply use small and
intermediate scale constraints on these parameters to predict the
large-scale $b_g$ for a given cosmology.

Increasing $\sigma_8$ boosts both $w_p$ and $\gamma_t$, but the impact
on $w_p$ is smaller because of the cancellation with decreased galaxy
bias at fixed $n_{\text{gal}}$. Raising $\sigma_8$ shifts the
inflection of $\xi_{\text{mm}}(r)$ at the 1-halo to 2-halo transition
outwards, because the virial radii of $M_{\star}$ halos are slightly
larger, which causes the jump in $d \ln w_p / d \ln \sigma_8$ at the
transition scale $r_p \approx 1-2 \, h^{-1}$ Mpc. The corresponding
effect in $\gamma_t(r_p)$ is a bump in the derivative at somewhat
larger scales.

Increasing $\Omega_m$ with fixed $\Omega_m h^2$ makes the matter power
spectrum bluer in observable, $h^{-1}$ Mpc units, decreasing $w_p$ at
large scales for fixed $\sigma_8$.\footnote{The power spectrum shape
  parameter $\Gamma = \Omega_m h$ increases, shifting the turnover in
  $P(k)$ to higher $k$ in $h$ Mpc$^{-1}$ units.} With linear evolution and linear
bias, there would be a compensating boost to $w_p$ at small scales,
but in our nonlinear calculation $w_p$ is suppressed at all $r_p$. By
contrast, increasing $\Omega_m$ boosts $\gamma_t(r_p)$ because of the
$\bar \rho$ factor in equation \ref{eq:ds}, but the logarithmic derivative is
below one because of the reduction in $\xi_{\text{mm}}(r)$ and the
increase of $\Sigma_c$ in equation \ref{eq:gamma_t}.

\begin{figure*}
  \includegraphics[width=\textwidth]{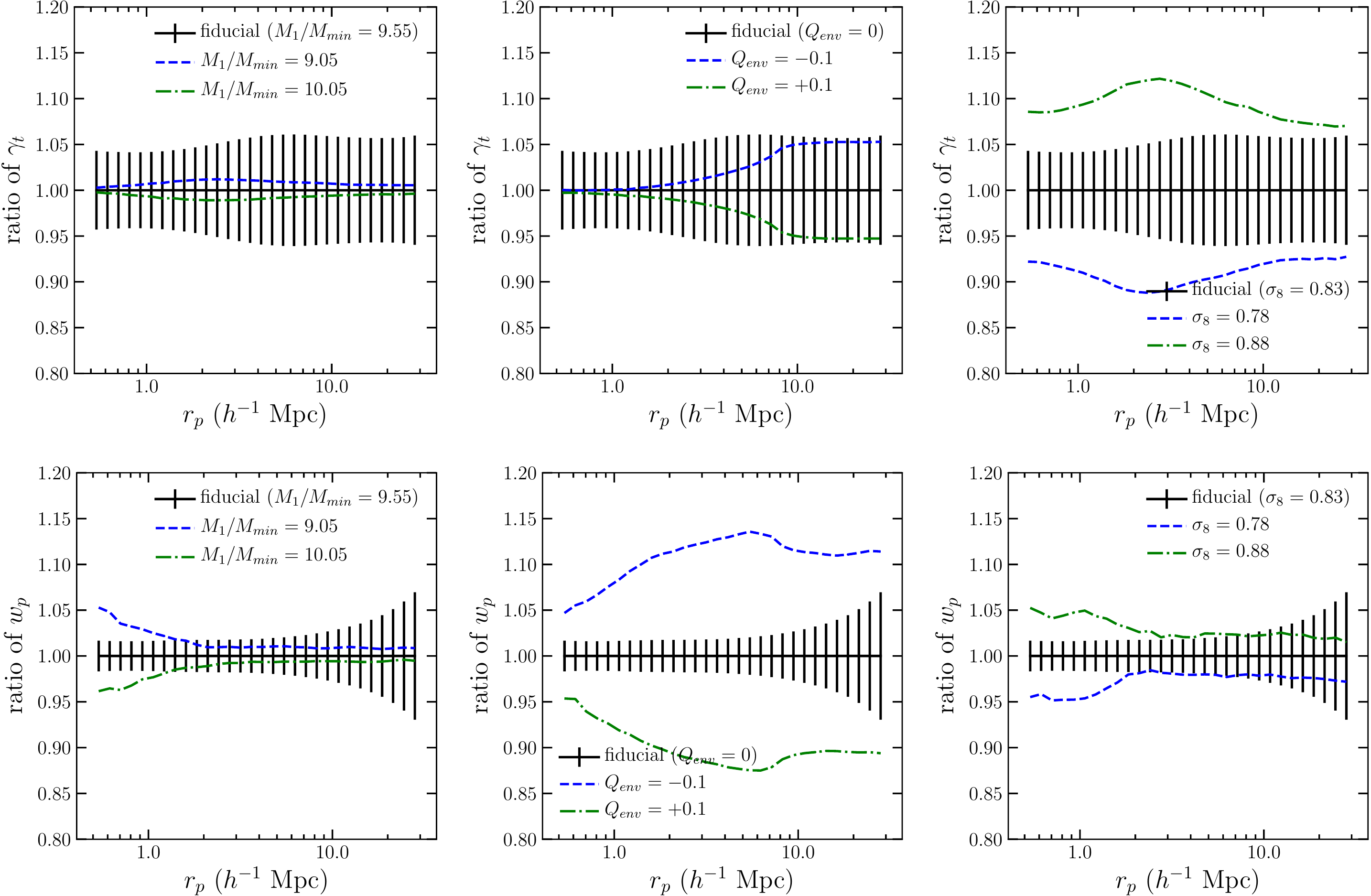}
  \caption{The ratios of observables for changes in some 
    parameters. The leftmost panel shows the effect of the HOD 
    parameter $M_1 / M_{\text{min}}$ on the observables $w_p$ and 
$\gamma_t$. The middle panel shows the effect of the environmental 
density parameter $Q_{\text{env}}$, and the rightmost panel shows the effect of 
$\sigma_8$. The error bars are the same as those shown in Figure \ref{fig:fiducial_emulator}.}
  \label{fig:ratios}
\end{figure*}

For a more concrete illustration of parameter impacts, Figure \ref{fig:ratios} shows
the fractional changes to $\gamma_t(r_p)$ and $w_p(r_p)$ that arise
from changing $M_1/M_{\text{min}}$ by $\pm 0.50$ from the fiducial
value of 9.55, changing $Q_{\text{env}}$ by $\pm 0.1$ from the
fiducial value of 0.0, or changing $\sigma_8$ by $\pm 0.05$ from the
fiducial value of 0.83. Here we have computed $\xi_{\text{gg}}(r)$ and
$\xi_{\text{gm}}(r)$ directly from the populated simulations, but
because these correspond to the same finite difference step sizes we use to compute the
$b_{\text{nl}}$, $b_g$, and $r_{\text{gm}}$ derivatives, the results
from using our emulator would be identical.

Changing $M_1/M_{\text{min}}$ alters the large scale amplitude of
$w_p(r_p)$ and $\gamma_t(r_p)$, and the impact grows at small scales
in $w_p(r_p)$ and intermediate scales in $\gamma_t(r_p)$. The effect
of $Q_{\text{env}}$, by contrast, is largest at large scales,
decreasing to nearly zero at sub-Mpc scales in $\gamma_t$. Raising or
lowering $\sigma_8$ raises or lowers the large scale $w_p(r_p)$ and
$\gamma_t(r_p)$ as expected, and non-linear evolution induces a
distinctive scale-dependence on scales of a few $h^{-1}$ Mpc and
below. The fact that each parameter produces a different
scale-dependence and has different effects on the two observables
demonstrates the potential of precise measurements across the full
range of scales to break degeneracies between cosmological quantities
and `nuisance' parameters that describe the relation between galaxies
and dark matter.

\subsection{Information and Forecasts}
\begin{figure*}
  \includegraphics[width=\textwidth]{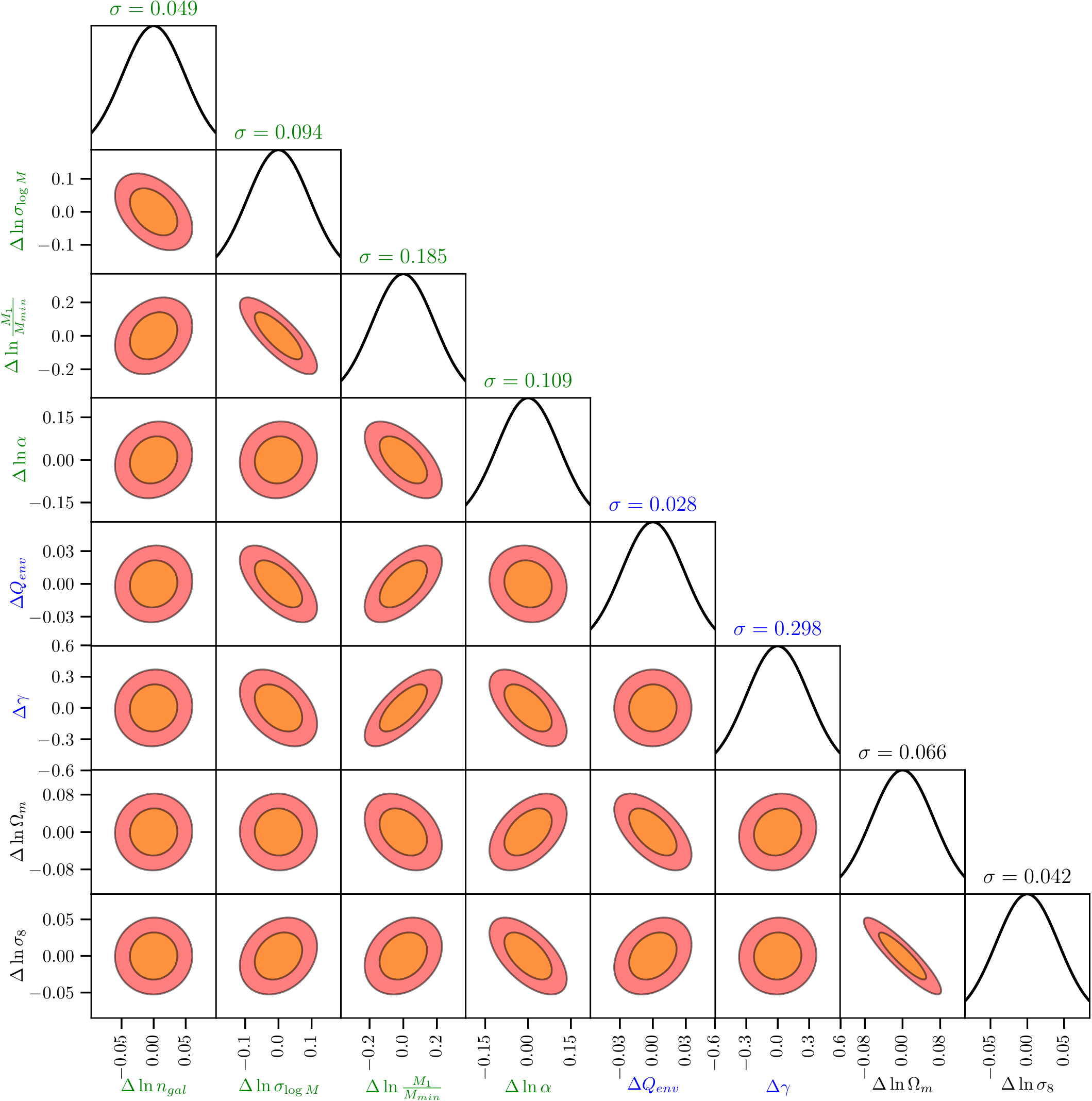}
  \caption{The fiducial forecast. We use the predicted projected
    galaxy correlation $w_p$ and the mean tangential shear $\gamma_t$ on scales $0.5 < r_p < 30$ $h^{-1}$ Mpc. Tight
    cosmological constraints are possible (4.2\% on $\sigma_8$, 6.6\%
    on $\Omega_m$) even after marginalizing over assembly bias and HOD
  parameters.}
\label{fig:ellipses}
\end{figure*}

Figure \ref{fig:ellipses} shows the parameter constraint forecasts for our fiducial
scenario, which adopts the $w_p$ and $\gamma_t$ covariance matrices of
Figure \ref{fig:cov} and a Gaussian prior on $\ln n_{\text{gal}}$ with a width of
5 per cent. All of the forecast parameters are in terms of the 
natural logarithm of the usual parameter, except for parameters that 
may plausibly be zero or negative (i.e., $Q_{\text{env}}$ and $\Delta 
\gamma$). With analysis down to scales of $0.5 \, h^{-1}$ Mpc, a data
combination like BOSS LOWZ and SDSS imaging can already yield
impressively tight constraints. The best-constrained combination of
cosmological parameters is $\sigma_8 \Omega_m^p$ with $p=0.58$, the
forecast uncertainty is 1.8 per cent after marginalizing over all HOD
parameters. The fully marginalized constraints on $\sigma_8$ and
$\Omega_m$ individually are 4.2 per cent and 6.6 per cent,
respectively.

The observational uncertainty in $n_{\text{gal}}$ will reflect both
cosmic variance and systematic uncertainties in completeness and
evolutionary corrections. Here we are treating our galaxy sample as
volume-limited and characterized by a single space density, but a full
observational analysis might require a redshift-dependent $\bar
n(z)$. For the individual luminosity-threshold samples of BOSS CMASS
galaxies studied by \cite{Guo_2014}, jackknife error estimates imply
$n_{\text{gal}}$ uncertainties of about 6 per cent (H. Guo, private communication).
We find that the forecast constraint on $n_{\text{gal}}$ is
essentially equal to our adopted prior (4.9 per cent vs. 5 per
cent). Fortunately, varying the $n_{\text{gal}}$ prior has negligible
impact on the cosmological parameter uncertainties; sharpening the
prior to 1 per cent or loosening it to 10 per cent does not change the
uncertainties in $\sigma_8$, $\Omega_m$, or $\sigma_8 \Omega_m^p$ at
the two-decimal-place precision quoted in our tables below.

Of other HOD parameters, the most poorly constrained is $\Delta
\gamma$, because its largest effects are limited to scales below the
smallest $r_p$ we consider. For the same reason, uncertainties in
$\Delta \gamma$ have little impact on the uncertainties in cosmological or other
HOD parameters. Uncertainties in $\sigma_{\log M}$,
$M_1/M_{\text{min}}$, and $\alpha$ are 9 per cent, 19 per cent and 11
per cent, respectively. Interestingly, the assembly bias parameter
$Q_{\text{env}}$ is quite tightly constrained, with a forecast
uncertainty of 0.028 dex. Changing the $n_{\text{gal}}$ prior to 10 per cent (1 per
cent) moderately loosens (tightens) the constraints on $\sigma_{\log
  M}$ and $M_1/M_{\text{min}}$ but has negligible effect on other
parameters.

Tables \ref{table:forecasts} and \ref{table:constrained} compare forecasts for a variety of other scenarios,
with Table \ref{table:forecasts} listing the marginalized constraints on $\sigma_8$,
$\Omega_m$, and individual HOD parameters and Table \ref{table:constrained} listing the best
constrained parameter combination of the form $\sigma_8
\Omega_m^p$. We first consider the impact of increasing the effective weak
lensing source density by a factor of 10 to 3 galaxies arcmin$^{-2}$, comparable to the
source density in the Dark Energy Survey instead of SDSS imaging. This
change lowers the shape noise contribution to the $\gamma_t$
covariance matrix (eq. \ref{eq:DS_cov}). The precision of $\sigma_8 \Omega_m^p$
improves by a factor of two, to 0.9 per cent. The individual
constraints on $\sigma_8$ and $\Omega_m$ improve by a factor $\approx
1.5$.

Returning to the fiducial source density of 1 arcmin$^{-2}$, we next
consider the impact of eliminating the $\gamma_t$ measurements at $r_p
< 5 \, h^{-1}$ Mpc. The constraint on $\sigma_8 \Omega_m^p$ degrades
to 2.2 per cent; the value of $p$ in the best constrained combination
depends on the data being considered, increasing slightly to $p=0.65$
in this case. Degradation can arise from the loss of aggregate
statistical precision in the $\gamma_t(r_p)$ measurement -- with fewer
points, the overall amplitude is less well determined and from the loss of leverage on parameter degeneracies for the reduced range of
scales. To isolate the second effect, we rescale the $\gamma_t(r_p)$
covariance matrix $C_{ij}$ by a constant factor that restores the
signal-to-noise ratio $(S/N) = \left[ D^{T} C^{-1} D \right]^{1/2}$ to its value in
the fiducial forecast, where $D$ is the data vector. With this rescaling, the constraint on $\sigma_8
\Omega_m^p$ \emph{improves} relative to the fiducial forecast, from
1.8 per cent to 1.2 per cent. Of course, one is not able to make this
adjustment in a real observational situation; the weak lensing error
bars at large scales do not decrease because one chooses to ignore
small scales in the modeling. However, this experiment shows that the
`per unit' information content of the large-scale $\gamma_t(r_p)$
measurements is more significant than the small scale measurements
because they suffer less degeneracy with galaxy bias parameters.  The
error bars on some HOD parameters, particularly $\sigma_{\log M}$,
$M_1/M_{\text{min}}$, and $Q_{\text{env}}$, do get worse when
eliminating small scale $\gamma_t(r_p)$ and rescaling the covariance
matrix.

For $w_p(r_p)$, the situation is reversed. Excluding points with $r_p
< 5 \, h^{-1}$ Mpc degrades the precision on $\sigma_8 \Omega_m^p$ by
more than a factor of two, from 1.8 per cent to 4.0 per cent, and
rescaling to restore the signal-to-noise ratio of the fiducial
measurement only improves the precision to 2.4 per cent. Without
rescaling, the constraints on HOD parameters become dramatically
worse, especially for the parameters $\alpha$ and $\Delta \gamma$
whose largest impact is on small scales. With rescaling, the
constraints on $\sigma_{\log M}$, $M_1/M_{\text{min}}$, and
$Q_{\text{env}}$ are actually better than in the fiducial case, but
the errors on $\alpha$ and $\Delta \gamma$ remain large, and the
degeneracy with $\sigma_8 \Omega_m^p$ is evidently large enough to
degrade its precision. The marginalized error on $\Omega_m$ itself
improves by nearly a factor of three over the fiducial case because of
the better measurement of the large scale shape of $w_p(r_p)$, and the
marginalized constraint on $\sigma_8$ improves moderately as a
result. Overall, this experiment shows that the information from
nonlinear scales of $w_p(r_p)$ improves the cosmological constraining
power of clustering and GGL by breaking degeneracies with HOD
parameters that describe the relation between galaxies and dark
matter. We caution that our Taylor expansion and Fisher matrix
calculation may become inaccurate with large uncertainties in $\alpha$
and $\Delta \gamma$, but this inaccuracy will not affect our
qualitative conclusions.

Finally, we consider the specific cuts adopted by
\cite{Mandelbaum_2013}, excluding $r_p < 2 \, h^{-1}$ Mpc for
$\gamma_t(r_p)$ and $r_p < 4 \, h^{-1}$ Mpc for $w_p(r_p)$. These cuts
increase the error on $\sigma_8 \Omega_m^p$ by more than a factor of
two, from 1.8 per cent to 3.8 per cent. The individual marginalized
constraints on $\sigma_8$ and $\Omega_m$ grow by slightly more and
slightly less than a factor of two, respectively. We conclude that the
gains in cosmological precision achievable from our more comprehensive
theoretical modeling, relative to the more conservative approach of
\cite{Mandelbaum_2013}, are about a factor of two in parameter errors,
equivalent to the effect of a fourfold increase in survey area.

\begin{figure}
  \includegraphics[width=\columnwidth]{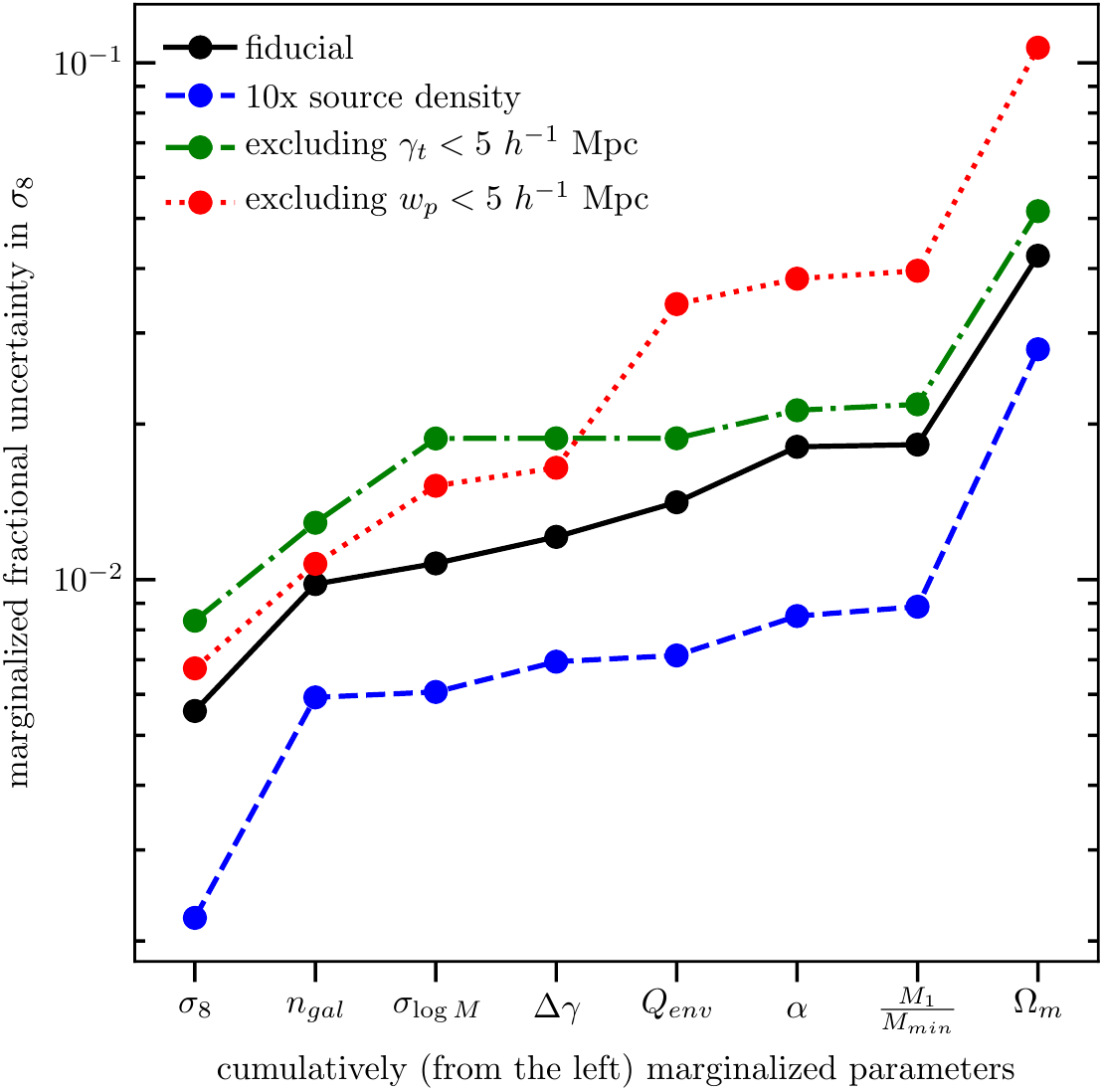}
  \caption{A comparison of forecasts on marginalized $\sigma_8$ for
    various scenarios. We show the marginalized fractional uncertainty
  on $\sigma_8$ as we marginalize over an increasing number of
  parameters from left-to-right. The order of parameters is set such
  that $\Omega_m$ is chosen to be the last-marginalized parameter,
  $\sigma_8$ is chosen to be the first parameter, and
the other parameters are ordered such that the steepest rise in
fractional uncertainty is obtained for the fiducial forecast
scenario.}
\label{fig:cumulative}
\end{figure}

Figure \ref{fig:cumulative} provides further insight into the degeneracy of cosmological
and HOD parameters and the role of $w_p(r_p)$ and $\gamma_t(r_p)$ in
breaking them. The leftmost points on each sequence show the
fractional uncertainty in $\sigma_8$ if all HOD parameters and
$\Omega_m$ are held fixed to their true values. Our fiducial data
combination could measure $\sigma_8$ to 0.56 per cent if all other
parameters were known perfectly. We then unleash the HOD parameters in
sequence, adding successively more degrees of freedom to the HOD
model. At each step in the sequence, we choose the parameter that
produces the sharpest increase of the $\sigma_8$ uncertainty when it
is set free, for the fiducial data case. Ordered this way, the
parameter with the highest leverage is $n_{\text{gal}}$, because with
other HOD parameters fixed a 5 per cent uncertainty in
$n_{\text{gal}}$ (set by our prior) can change the galaxy bias factor
significantly. We previously found that varying the $n_{\text{gal}}$
prior from 0.01 to 0.10 had negligible impact on cosmological
precision (see Tables \ref{table:forecasts} and \ref{table:constrained}), but that was with other HOD parameters free to compensate
for its effect. Adding more HOD parameters steadily increases the
marginalized $\sigma_8$ error, reaching 1.8 per cent with the full
parameter set. Because of the usual cosmological degeneracy between
$\sigma_8$ and $\Omega_m$, the marginalized $\sigma_8$ error rises to
4.2 per cent when $\Omega_m$ is also free. However, the fractional error on
$\sigma_8$ with fixed $\Omega_m$ is the same as the fractional error
on the best constrained $\sigma_8 \Omega_m^p$ combination, and we view
this as the best characterization of the statistical power of a
combined clustering + GGL data set.

When small scales of $\gamma_t(r_p)$ are dropped (green curve in Figure
\ref{fig:cumulative}), the precision on $\sigma_8$ as the sole free parameter degrades by
a factor of 1.5, from 0.56 per cent to 0.83 per cent. However, the uncertainty associated with scale-dependent
galaxy bias is reduced when the small scale $\gamma_t(r_p)$ are not
considered, so the degradation with all HOD parameters free is only a
factor of 1.2 (2.2 per cent versus 1.8 per cent, as listed in Table \ref{table:constrained}).

Dropping the small scale $w_p(r_p)$ data instead (red curve) produces
minimal degradation when HOD parameters are fixed, but now these
parameters are poorly constrained and thus have a large impact once
they are set free. In particular, the assembly bias parameter
$Q_{\text{env}}$ has a much larger uncertainty in this case (see Table
\ref{table:forecasts}) and has a more pronounced impact on cosmological parameter
uncertainty. Raising the weak lensing source density (blue curve)
improves the $\sigma_8$ precision at fixed HOD by a factor of 2.5
(0.22 per cent versus 0.56 per cent), nearly the full factor of
$10^{1/3} = 3.16$ that would be expected if weak lensing shape noise
were the only effect limiting the measurement precision. The relative
impact of galaxy bias uncertainties is larger when the weak lensing
precision is higher, but not drastically so; when all HOD parameters
are free, the $\sigma_8$ precision is still a factor of two better
than that of the fiducial case (0.9 per cent versus 1.8 per cent).

The final lines of Tables \ref{table:forecasts} and \ref{table:constrained} show forecasts based on the
$w_p(r_p)$ data alone, with no GGL information. In a pure linear
theory calculation with galaxy bias $b_g$ as a free parameter, the
shape of $w_p(r_p)$ would constrain $\Omega_m$, but there would be no
constraint on $\sigma_8$ because it would be fully degenerate with
$b_g$. Our non-linear forecast with an HOD description of galaxy bias
yields an 11 per cent constraint on $\sigma_8 \Omega_m^p$ and a 12 per
cent marginalized constraint on $\sigma_8$. These are much worse than
the 1.8 per cent and 4.2 per cent fiducial forecasts, demonstrating
that the great majority of the cosmological information is coming from
the \emph{combination} of clustering and GGL, not from the high
precision clustering measurements on their own. The uncertainty in
$Q_{\text{env}}$ is also substantially larger for the clustering only
case (0.058 vs. 0.028). This difference shows that, while some of the
information about $Q_{\text{env}}$ is coming from the distinctive scale
dependence that it produces, much of it coming from the relative
strength of clustering and GGL. This suggests that the clustering +
GGL combination could be a useful diagnostic of galaxy assembly bias,
especially if one has strong external constraints on
$\Omega_m$. Physically, assembly bias is the main effect that can alter
the large scale galaxy bias given constraints on other HOD parameters
from small and intermediate scales. The
clustering + GGL combination breaks the degeneracy of $b_g$ and
$\sigma_8$, even on linear scales, so it can test for the presence or absence of this effect.
\begin{table*}
\caption{Forecasts of fractional uncertainties in cosmological and HOD 
  parameters for various scenarios. For some scenarios, we show the 
  forecast where the signal-to-noise is rescaled to match the 
  signal-to-noise of the fiducial forecast (indicated by ``rescaled
  S/N'').}
\label{table:forecasts}
\begin{tabular}{lcccccccc}
\toprule
{} & $\Delta \ln n_{\textrm{gal}}$ & $\Delta \ln \sigma_{\log M}$ & $\Delta \ln \frac{M_1}{M_{\textrm{min}}}$ & $\Delta \ln \alpha$ & $\Delta Q_{\textrm{env}}$ & $\Delta \gamma$ & $\Delta \ln \Omega_m$ & $\Delta \ln \sigma_8$ \\
\midrule
fiducial                                                    & 0.049 & 0.094 & 0.185 & 0.109 & 0.028 & 0.298 & 0.066 & 0.042 \\
10x source density                                          & 0.048 & 0.067 & 0.165 & 0.104 & 0.017 & 0.264 & 0.045 & 0.028 \\
excluding $\gamma_t < 5$ $h^{-1}$ Mpc                       & 0.049 & 0.160 & 0.239 & 0.112 & 0.049 & 0.318 & 0.072 & 0.052 \\
excluding $\gamma_t < 5$ $h^{-1}$ Mpc (rescaled S/N)        & 0.049 & 0.141 & 0.218 & 0.109 & 0.036 & 0.298 & 0.065 & 0.047 \\
excluding $w_p < 5$ $h^{-1}$ Mpc                            & 0.050 & 0.290 & 0.454 & 1.072 & 0.066 & 1.726 & 0.113 & 0.107 \\
excluding $w_p < 5$ $h^{-1}$ Mpc (rescaled S/N)             & 0.049 & 0.078 & 0.131 & 0.429 & 0.024 & 0.574 & 0.024 & 0.034 \\
excluding both $< 5 \, h^{-1}$ Mpc                          & 0.050 & 0.313 & 0.617 & 1.944 & 0.103 & 21.130 & 0.124 & 0.119 \\
excluding $< 2$ ($\gamma_t$) and $< 4$ ($w_p$) $h^{-1}$ Mpc & 0.050 & 0.265 & 0.513 & 1.697 & 0.084 & 4.057 & 0.105 & 0.094 \\
$\Delta \ln n_{\text{gal}} = 0.01$                          & 0.010 & 0.085 & 0.179 & 0.108 & 0.028 & 0.297 & 0.066 & 0.042 \\
$\Delta \ln n_{\text{gal}} = 0.1$                           & 0.091 & 0.113 & 0.201 & 0.111 & 0.029 & 0.299 & 0.066 & 0.042 \\
no lensing                                                  & 0.050 & 0.189 & 0.320 & 0.151 & 0.058 & 0.340 & 0.113 & 0.118 \\
\bottomrule
\end{tabular}

\end{table*}
\begin{table*}
\caption{Best-constrained parameters for forecasts.}
\label{table:constrained}
\begin{tabular}{lcc}
\toprule
{} &    $p$ & best-constrained $\sigma_8 \Omega_m^p$ \\
\midrule
fiducial                                                    & 0.576 & 0.018 \\
10x source density                                          & 0.584 & 0.009 \\
excluding $\gamma_t < 5$ $h^{-1}$ Mpc                       & 0.646 & 0.022 \\
excluding $\gamma_t < 5$ $h^{-1}$ Mpc (rescaled S/N)        & 0.707 & 0.012 \\
excluding $w_p < 5$ $h^{-1}$ Mpc                            & 0.881 & 0.040 \\
excluding $w_p < 5$ $h^{-1}$ Mpc (rescaled S/N)             & 0.999 & 0.024 \\
excluding both $< 5 \, h^{-1}$ Mpc                          & 0.885 & 0.046 \\
excluding $< 2$ ($\gamma_t$) and $< 4$ ($w_p$) $h^{-1}$ Mpc & 0.813 & 0.038 \\
$\Delta \ln n_{\text{gal}} = 0.01$                          & 0.576 & 0.018 \\
$\Delta \ln n_{\text{gal}} = 0.1$                           & 0.575 & 0.018 \\
no lensing                                                  & -0.432 & 0.107 \\
\bottomrule
\end{tabular}

\end{table*}

\section{Discussion and Conclusions}
\label{section:discussion}
Several observational studies have demonstrated the promise of
combining galaxy clustering and GGL to constrain cosmological
parameters and test $\Lambda$CDM+GR predictions of matter clustering
\citep{Mandelbaum_2013,More_2015,Hildebrandt_2017,Leauthaud_2017,Prat_2017,DESY1KP}.
Building on earlier theoretical studies by \cite{Yoo_2006}, \cite{Leauthaud_2011},
\cite{Cacciato_2012}, and \cite{Yoo_2012}, our investigation
demonstrates the power of extending these analyses down to 
small scales using a flexible model for the relation between
galaxies and dark matter.  As a fiducial case, we consider 
HOD parameters and covariance matrices scaled to the 
BOSS LOWZ galaxy lens sample and SDSS-depth imaging
(approximately one source galaxy per arcmin$^2$) over 
9,000 deg$^2$ \citep{Singh_2016} for weak lensing measurements.
Extending the analysis of $\gamma_t(r_p)$ and $w_p(r_p)$
to $0.5 \, h^{-1}$ Mpc improves the precision of the best-constrained
$\sigma_8\Omega_m^p$ combination by more than a factor of two
(1.8\% vs. 3.8\%) relative to the more conservative
cuts ($2 \, h^{-1}$ Mpc for $\gamma_t$ and $4 \, h^{-1}$ Mpc for $w_p$)
adopted by \cite{Mandelbaum_2013} for their analysis of SDSS DR7,
which uses the perturbative bias model of \cite{Baldauf_2010}.
Some of the gain in parameter precision comes directly from
using the small scale $\gamma_t(r_p)$ measurements, which
provide additional leverage on the amplitude of the galaxy-matter
cross-correlation.  However, the largest gains come from using
the smaller scales of $w_p(r_p)$ to constrain HOD parameters,
which allows our model to make better use of the large scale
$\gamma_t(r_p)$ data for the cosmological constraints.

The emulator approach described in \S\ref{section:emulator} makes a
fully non-linear, $N$-body + HOD approach practical for statistical
analysis.  In this paper we have considered only $\sigma_8$ and
$\Omega_m$ as the varying cosmological parameters, and we have
computed results at $z = 0.3$.  However, because we construct
our emulator to compute ratios of correlation functions 
starting from $\xi_{\rm mm,lin}$, it can accommodate some
range of cosmological parameters.  We will improve the emulator
in future work using a grid of cosmological simulations, which
will also allow leave-one-out tests for the emulator's accuracy
and more systematic study of its range of validity.
Detailed predictions for an observational data set also require
information about the redshift distributions of the lens and source
samples and may include nuisance parameters that describe
observational or theoretical systematics.  Rather than incorporate
these survey-specific elements into our emulator, we focus on
predicting the two quantities, $\xi_{\text{gg}}$ and $\xi_{\text{gm}}$,
that require non-linear clustering calculations.

The main limitation of the emulator approach as pursued here is the
need to retune the fiducial HOD for each galaxy lens sample, and to
model the range of redshifts probed by that sample.  This does not
require new simulations, but it does require new HOD populations
and correlation function measurements for each lens sample 
being considered.

A novel aspect of our model is inclusion of a parameterized description
of HOD environmental variation, to capture the potential effects of
galaxy assembly bias.  This prescription allows the large scale 
galaxy bias to be at least partly decoupled from the ``classic''
HOD parameters constrained by small and intermediate scale clustering.
It is encouraging that this new degree of freedom in the galaxy bias
model does not lead to substantial degradation of the cosmological
parameter constraints.  Indeed, we find that the combination of 
clustering and GGL gives interestingly tight constraints on $Q_{\text{env}}$
even with free cosmological parameters.  To date, most observational
tests for galaxy assembly bias have focused on comparing clustering
of blue and red galaxies, but the approach outlined here
could provide a way to test for assembly bias in luminosity- or
mass-selected galaxy samples.  Our $Q_{\text{env}}$ parameterization predicts
a scale-dependence of galaxy bias that might be different from that 
predicted by a specific physical model that ties galaxy properties
to halo assembly.  The best way to test the adequacy of our model is 
to apply it to galaxy populations drawn from hydrodynamic simulations
or to simulations that populate $N$-body halos using abundance and
age-matching prescriptions (e.g., \citealt{Hearin_2013,Lehmann_2017})
or semi-analytic galaxy formation models.  Hydrodynamic simulations
are also needed to test for baryonic effects on the mass distribution,
including the impact of subhalos around satellite
galaxies. \cite{Yoo_2006} found little impact of subhalos on GGL, but
their hydrodynamic simulations were too small for tests at the level
of precision needed for current data sets.

The stakes for precise and accurate joint clustering and GGL analyses are high,
because many cosmic shear and GGL analyses to date yield estimates
of $\sigma_8\Omega_m^{0.5}$ that are lower than that predicted by
a Planck-normalized $\Lambda$CDM model (e.g. \citealt{Heymans_2012,Hildebrandt_2017,DESY1KP}).  The statistical
significance for any one data set is usually $\la 2\sigma$. Improving lensing and CMB analyses could remove this tension, or
they could sharpen it into strong evidence for new physics.
The largest magnitude of discrepancy is found by \cite{Leauthaud_2017},
who measure GGL for BOSS CMASS galaxies from 250 deg$^2$ of deep
imaging data and compare their measurements to predictions from
a variety of mock catalogs that are designed to reproduce observed
CMASS galaxy clustering.  On scales of $0.3-3 \, h^{-1}$ Mpc the discrepancy
in $\gamma_t(r_p)$ is about 20\% and well outside the statistical
errors, and on scales of $0.1-0.3 \, h^{-1}$ Mpc it is larger still.
The results of \cite{McEwen_2016} suggest that this discrepancy should
be robust to uncertainties about galaxy assembly bias, because they
find that even when strong assembly bias is present,
an HOD model that reproduces the observed galaxy clustering also 
predicts the correct ratio $\xi_{\text{gm}}(r)/\xi_{\text{gg}}(r)$.
Our results here provide further support for this view, showing
that clustering and GGL can yield strong cosmological constraints
even when marginalizing over our parameterized assembly bias prescription.

For our fiducial data assumptions, we forecast a 1.8\% error on
$\sigma_8\Omega_m^{0.58}$.  For comparison, the error on
$S_8 \equiv \sigma_8(\Omega_m/0.3)^{0.5}$ from DES Year 1 data
is 2.9\% \citep{DESY1KP}.  
This constraint uses clustering \citep{ElvinPoole_2017} and GGL \citep{Prat_2017} of the DES ``redMaGiC'' 
galaxy sample \citep{ElvinPoole_2017,Rozo_2016} {\it and} cosmic shear from the 
same imaging data \citep{Troxel_2017}, comprising 26 million galaxy
shape measurements over 1321 deg$^2$ (5.4 gal/arcmin$^2$).
The DES analysis includes marginalization over several systematics
not considered here, such as shear calibration uncertainties,
photometric redshift biases, and galaxy intrinsic alignments.
These uncertainties must be accounted for in any GGL analysis,
and they will degrade the precision of cosmological measurements
below that of our forecasts.  Nonetheless, our results show that
extending to non-linear scales allows even SDSS-depth imaging to
achieve constraints competitive with the best current weak lensing
data sets.

Our emulator results (detailed in Appendix \ref{appendix:derivatives}) can be applied as
they are to GGL measurements of the BOSS LOWZ sample.  For
imaging data significantly deeper than SDSS, better constraints
will come from higher redshift lens populations that probe larger volumes,
such as the BOSS CMASS spectroscopic sample or photometrically
defined samples such as DES redMaGiC.  We will investigate predictions
for such samples in future work.  There are many considerations
that go into choosing a lens sample, including lens density and
redshift distribution, overlap with deep imaging data for spectroscopic
samples, accuracy of photometric redshifts for photometric samples,
and observational uncertainties such as incompleteness, contamination,
or depth variations.  Our results suggest that physical simplicity
should be an additional consideration in defining lens samples,
since the extension of analyses to non-linear scales can substantially
improve their constraining power but requires accurate modeling. Data
sets emerging over the next few years should enable tests of the
matter clustering predicted by General Relativity at the percent or
even sub-percent level, with the potential to reveal profound new
physics or to provide powerful confirmation of the reigning theories
of dark energy and cosmological gravity.

\section*{Acknowledgements}
We thank Chris Hirata, Eric Huff, Rachel Mandelbaum, Sukhdeep Singh,
and Ying Zu for valuable conversations about this work.

BDW is supported by the
National Science Foundation Graduate Research Fellowship Program under
Grant No. DGE-1343012. ANS is supported by the Department of Energy
Computational Science Graduate Fellowship Program of the Office of
Science and National Nuclear Security Administration in the Department
of Energy under contract DE-FG02-97ER25308.  BDW, ANS, and DHW are supported
in part by NSF grant AST-1516997. LG and DJE have been supported by
National Science Foundation grant AST-1313285. DJE is further
supported by DOE-SC0013718 and as a Simons Foundation Investigator. Any opinions, 
findings, and conclusions or recommendations expressed in this material are those of the 
author(s) and do not necessarily reflect the views of the National 
Science Foundation. 

Simulations were analyzed in part on computational resources of the
Ohio Supercomputer Center \citep{OhioSupercomputerCenter1987},
with resources supported in part by the Center for Cosmology and AstroParticle
Physics at the Ohio State University. Some computations in this paper were performed on the El Gato supercomputer at the University of Arizona, supported by grant 1228509 from the National Science Foundation, and on the Odyssey cluster supported by the FAS Division of Science, Research Computing Group at Harvard University.

We gratefully acknowledge the use of the \textsc{matplotlib}
\citep{Hunter_2007} software package and the GNU Scientific Library
\citep{GSL_2009}. This research has made use of NASA's Astrophysics Data System.



\bibliographystyle{mnras}
\bibliography{bibliography/biblio}

\begin{thebibliography}{}
\makeatletter
\relax
\def\mn@urlcharsother{\let\do\@makeother \do\$\do\&\do\#\do\^\do\_\do\%\do\~}
\def\mn@doi{\begingroup\mn@urlcharsother \@ifnextchar [ {\mn@doi@}
  {\mn@doi@[]}}
\def\mn@doi@[#1]#2{\def\@tempa{#1}\ifx\@tempa\@empty \href
  {http://dx.doi.org/#2} {doi:#2}\else \href {http://dx.doi.org/#2} {#1}\fi
  \endgroup}
\def\mn@eprint#1#2{\mn@eprint@#1:#2::\@nil}
\def\mn@eprint@arXiv#1{\href {http://arxiv.org/abs/#1} {{\tt arXiv:#1}}}
\def\mn@eprint@dblp#1{\href {http://dblp.uni-trier.de/rec/bibtex/#1.xml}
  {dblp:#1}}
\def\mn@eprint@#1:#2:#3:#4\@nil{\def\@tempa {#1}\def\@tempb {#2}\def\@tempc
  {#3}\ifx \@tempc \@empty \let \@tempc \@tempb \let \@tempb \@tempa \fi \ifx
  \@tempb \@empty \def\@tempb {arXiv}\fi \@ifundefined
  {mn@eprint@\@tempb}{\@tempb:\@tempc}{\expandafter \expandafter \csname
  mn@eprint@\@tempb\endcsname \expandafter{\@tempc}}}

\bibitem[\protect\citeauthoryear{{Abazajian} et~al.,}{{Abazajian}
  et~al.}{2009}]{Abazajian_2009}
{Abazajian} K.~N.,  et~al., 2009, \mn@doi [\apjs]
  {10.1088/0067-0049/182/2/543}, \href
  {http://adsabs.harvard.edu/abs/2009ApJS..182..543A} {182, 543}

\bibitem[\protect\citeauthoryear{{Aihara} et~al.,}{{Aihara}
  et~al.}{2017}]{HSC2017}
{Aihara} H.,  et~al., 2017, preprint, \href
  {http://ads.nao.ac.jp/abs/2017arXiv170405858A} {} (\mn@eprint {arXiv}
  {1704.05858})

\bibitem[\protect\citeauthoryear{{Baldauf}, {Smith}, {Seljak}  \&
  {Mandelbaum}}{{Baldauf} et~al.}{2010}]{Baldauf_2010}
{Baldauf} T.,  {Smith} R.~E.,  {Seljak} U.,   {Mandelbaum} R.,  2010, \mn@doi
  [\prd] {10.1103/PhysRevD.81.063531}, \href
  {http://adsabs.harvard.edu/abs/2010PhRvD..81f3531B} {81, 063531}

\bibitem[\protect\citeauthoryear{{Behroozi}, {Wechsler}  \& {Wu}}{{Behroozi}
  et~al.}{2013}]{Behroozi_2013}
{Behroozi} P.~S.,  {Wechsler} R.~H.,   {Wu} H.-Y.,  2013, \mn@doi [\apj]
  {10.1088/0004-637X/762/2/109}, \href
  {http://adsabs.harvard.edu/abs/2013ApJ...762..109B} {762, 109}

\bibitem[\protect\citeauthoryear{{Benson}, {Cole}, {Frenk}, {Baugh}  \&
  {Lacey}}{{Benson} et~al.}{2000}]{Benson_2000}
{Benson} A.~J.,  {Cole} S.,  {Frenk} C.~S.,  {Baugh} C.~M.,   {Lacey} C.~G.,
  2000, \mn@doi [\mnras] {10.1046/j.1365-8711.2000.03101.x}, \href
  {http://adsabs.harvard.edu/abs/2000MNRAS.311..793B} {311, 793}

\bibitem[\protect\citeauthoryear{{Berlind} \& {Weinberg}}{{Berlind} \&
  {Weinberg}}{2002}]{Berlind_2002}
{Berlind} A.~A.,  {Weinberg} D.~H.,  2002, \mn@doi [\apj] {10.1086/341469},
  \href {http://adsabs.harvard.edu/abs/2002ApJ...575..587B} {575, 587}

\bibitem[\protect\citeauthoryear{{Bond}, {Cole}, {Efstathiou}  \&
  {Kaiser}}{{Bond} et~al.}{1991}]{Bond_1991}
{Bond} J.~R.,  {Cole} S.,  {Efstathiou} G.,   {Kaiser} N.,  1991, \mn@doi
  [\apj] {10.1086/170520}, \href
  {http://adsabs.harvard.edu/abs/1991ApJ...379..440B} {379, 440}

\bibitem[\protect\citeauthoryear{{Bryan} \& {Norman}}{{Bryan} \&
  {Norman}}{1998}]{Bryan_1998}
{Bryan} G.~L.,  {Norman} M.~L.,  1998, \mn@doi [\apj] {10.1086/305262}, \href
  {http://adsabs.harvard.edu/abs/1998ApJ...495...80B} {495, 80}

\bibitem[\protect\citeauthoryear{{Cacciato}, {Lahav}, {van den Bosch},
  {Hoekstra}  \& {Dekel}}{{Cacciato} et~al.}{2012}]{Cacciato_2012}
{Cacciato} M.,  {Lahav} O.,  {van den Bosch} F.~C.,  {Hoekstra} H.,   {Dekel}
  A.,  2012, \mn@doi [\mnras] {10.1111/j.1365-2966.2012.21762.x}, \href
  {http://adsabs.harvard.edu/abs/2012MNRAS.426..566C} {426, 566}

\bibitem[\protect\citeauthoryear{Center}{Center}{1987}]{OhioSupercomputerCenter1987}
Center O.~S.,  1987, Ohio Supercomputer Center,
  \url{http://osc.edu/ark:/19495/f5s1ph73}

\bibitem[\protect\citeauthoryear{{Cooray} \& {Hu}}{{Cooray} \&
  {Hu}}{2001}]{Cooray_2001}
{Cooray} A.,  {Hu} W.,  2001, \mn@doi [\apj] {10.1086/321376}, \href
  {http://adsabs.harvard.edu/abs/2001ApJ...554...56C} {554, 56}

\bibitem[\protect\citeauthoryear{{Correa}, {Wyithe}, {Schaye}  \&
  {Duffy}}{{Correa} et~al.}{2015}]{Correa_2015}
{Correa} C.~A.,  {Wyithe} J.~S.~B.,  {Schaye} J.,   {Duffy} A.~R.,  2015,
  \mn@doi [\mnras] {10.1093/mnras/stv1363}, \href
  {http://adsabs.harvard.edu/abs/2015MNRAS.452.1217C} {452, 1217}

\bibitem[\protect\citeauthoryear{{Coupon} et~al.,}{{Coupon}
  et~al.}{2012}]{Coupon_2012}
{Coupon} J.,  et~al., 2012, \mn@doi [\aap] {10.1051/0004-6361/201117625}, \href
  {http://adsabs.harvard.edu/abs/2012A%26A...542A...5C} {542, A5}

\bibitem[\protect\citeauthoryear{{DES Collaboration} et~al.,}{{DES
  Collaboration} et~al.}{2017}]{DESY1KP}
{DES Collaboration} et~al., 2017, preprint, \href
  {http://adsabs.harvard.edu/abs/2017arXiv170801530D} {} (\mn@eprint {arXiv}
  {1708.01530})

\bibitem[\protect\citeauthoryear{{Dawson} et~al.,}{{Dawson}
  et~al.}{2013}]{Dawson_2013}
{Dawson} K.~S.,  et~al., 2013, \mn@doi [\aj] {10.1088/0004-6256/145/1/10},
  \href {http://adsabs.harvard.edu/abs/2013AJ....145...10D} {145, 10}

\bibitem[\protect\citeauthoryear{{Eisenstein} et~al.,}{{Eisenstein}
  et~al.}{2011}]{Eisenstein_2011}
{Eisenstein} D.~J.,  et~al., 2011, \mn@doi [\aj] {10.1088/0004-6256/142/3/72},
  \href {http://adsabs.harvard.edu/abs/2011AJ....142...72E} {142, 72}

\bibitem[\protect\citeauthoryear{{Elvin-Poole} et~al.,}{{Elvin-Poole}
  et~al.}{2017}]{ElvinPoole_2017}
{Elvin-Poole} J.,  et~al., 2017, preprint, \href
  {http://adsabs.harvard.edu/abs/2017arXiv170801536E} {} (\mn@eprint {arXiv}
  {1708.01536})

\bibitem[\protect\citeauthoryear{Galassi, Davies, Theiler, Gough, Jungman,
  Alken, Booth  \& Rossi}{Galassi et~al.}{2009}]{GSL_2009}
Galassi M.,  Davies J.,  Theiler J.,  Gough B.,  Jungman G.,  Alken P.,  Booth
  M.,   Rossi F.,  2009, GNU Scientific Library Reference Manual.
3 edn

\bibitem[\protect\citeauthoryear{{Gao}, {Springel}  \& {White}}{{Gao}
  et~al.}{2005}]{Gao_2005}
{Gao} L.,  {Springel} V.,   {White} S.~D.~M.,  2005, \mn@doi [\mnras]
  {10.1111/j.1745-3933.2005.00084.x}, \href
  {http://adsabs.harvard.edu/abs/2005MNRAS.363L..66G} {363, L66}

\bibitem[\protect\citeauthoryear{{Garrison}, {Eisenstein}, {Ferrer}, {Metchnik}
   \& {Pinto}}{{Garrison} et~al.}{2016}]{Garrison_2016}
{Garrison} L.~H.,  {Eisenstein} D.~J.,  {Ferrer} D.,  {Metchnik} M.~V.,
  {Pinto} P.~A.,  2016, \mn@doi [\mnras] {10.1093/mnras/stw1594}, \href
  {http://adsabs.harvard.edu/abs/2016MNRAS.461.4125G} {461, 4125}

\bibitem[\protect\citeauthoryear{{Garrison} et~al.}{{Garrison}
  et~al.}{2017}]{Garrison_2017}
{Garrison} L.,  et~al., {in prep, 2017.}

\bibitem[\protect\citeauthoryear{{Guo} et~al.,}{{Guo} et~al.}{2014}]{Guo_2014}
{Guo} H.,  et~al., 2014, \mn@doi [\mnras] {10.1093/mnras/stu763}, \href
  {http://adsabs.harvard.edu/abs/2014MNRAS.441.2398G} {441, 2398}

\bibitem[\protect\citeauthoryear{{Harker}, {Cole}, {Helly}, {Frenk}  \&
  {Jenkins}}{{Harker} et~al.}{2006}]{Harker_2006}
{Harker} G.,  {Cole} S.,  {Helly} J.,  {Frenk} C.,   {Jenkins} A.,  2006,
  \mn@doi [\mnras] {10.1111/j.1365-2966.2006.10022.x}, \href
  {http://adsabs.harvard.edu/abs/2006MNRAS.367.1039H} {367, 1039}

\bibitem[\protect\citeauthoryear{{Hearin} \& {Watson}}{{Hearin} \&
  {Watson}}{2013}]{Hearin_2013}
{Hearin} A.~P.,  {Watson} D.~F.,  2013, \mn@doi [\mnras]
  {10.1093/mnras/stt1374}, \href
  {http://adsabs.harvard.edu/abs/2013MNRAS.435.1313H} {435, 1313}

\bibitem[\protect\citeauthoryear{{Heitmann}, {Higdon}, {White}, {Habib},
  {Williams}, {Lawrence}  \& {Wagner}}{{Heitmann} et~al.}{2009}]{Heitmann_2009}
{Heitmann} K.,  {Higdon} D.,  {White} M.,  {Habib} S.,  {Williams} B.~J.,
  {Lawrence} E.,   {Wagner} C.,  2009, \mn@doi [\apj]
  {10.1088/0004-637X/705/1/156}, \href
  {http://adsabs.harvard.edu/abs/2009ApJ...705..156H} {705, 156}

\bibitem[\protect\citeauthoryear{{Heymans} et~al.,}{{Heymans}
  et~al.}{2012}]{Heymans_2012}
{Heymans} C.,  et~al., 2012, \mn@doi [\mnras]
  {10.1111/j.1365-2966.2012.21952.x}, \href
  {http://adsabs.harvard.edu/abs/2012MNRAS.427..146H} {427, 146}

\bibitem[\protect\citeauthoryear{{Hildebrandt} et~al.,}{{Hildebrandt}
  et~al.}{2017}]{Hildebrandt_2017}
{Hildebrandt} H.,  et~al., 2017, \mn@doi [\mnras] {10.1093/mnras/stw2805},
  \href {http://adsabs.harvard.edu/abs/2017MNRAS.465.1454H} {465, 1454}

\bibitem[\protect\citeauthoryear{Hunter}{Hunter}{2007}]{Hunter_2007}
Hunter J.~D.,  2007, \mn@doi [Computing In Science \& Engineering]
  {10.1109/MCSE.2007.55}, 9, 90

\bibitem[\protect\citeauthoryear{{Jain} \& {Seljak}}{{Jain} \&
  {Seljak}}{1997}]{Jain_1997}
{Jain} B.,  {Seljak} U.,  1997, \mn@doi [\apj] {10.1086/304372}, \href
  {http://adsabs.harvard.edu/abs/1997ApJ...484..560J} {484, 560}

\bibitem[\protect\citeauthoryear{{Jee}, {Tyson}, {Hilbert}, {Schneider},
  {Schmidt}  \& {Wittman}}{{Jee} et~al.}{2016}]{Jee_2016}
{Jee} M.~J.,  {Tyson} J.~A.,  {Hilbert} S.,  {Schneider} M.~D.,  {Schmidt} S.,
   {Wittman} D.,  2016, \mn@doi [\apj] {10.3847/0004-637X/824/2/77}, \href
  {http://adsabs.harvard.edu/abs/2016ApJ...824...77J} {824, 77}

\bibitem[\protect\citeauthoryear{{Leauthaud}, {Tinker}, {Behroozi}, {Busha}  \&
  {Wechsler}}{{Leauthaud} et~al.}{2011}]{Leauthaud_2011}
{Leauthaud} A.,  {Tinker} J.,  {Behroozi} P.~S.,  {Busha} M.~T.,   {Wechsler}
  R.~H.,  2011, \mn@doi [\apj] {10.1088/0004-637X/738/1/45}, \href
  {http://adsabs.harvard.edu/abs/2011ApJ...738...45L} {738, 45}

\bibitem[\protect\citeauthoryear{{Leauthaud} et~al.,}{{Leauthaud}
  et~al.}{2017}]{Leauthaud_2017}
{Leauthaud} A.,  et~al., 2017, \mn@doi [\mnras] {10.1093/mnras/stx258}, \href
  {http://adsabs.harvard.edu/abs/2017MNRAS.467.3024L} {467, 3024}

\bibitem[\protect\citeauthoryear{{Lehmann}, {Mao}, {Becker}, {Skillman}  \&
  {Wechsler}}{{Lehmann} et~al.}{2017}]{Lehmann_2017}
{Lehmann} B.~V.,  {Mao} Y.-Y.,  {Becker} M.~R.,  {Skillman} S.~W.,   {Wechsler}
  R.~H.,  2017, \mn@doi [\apj] {10.3847/1538-4357/834/1/37}, \href
  {http://adsabs.harvard.edu/abs/2017ApJ...834...37L} {834, 37}

\bibitem[\protect\citeauthoryear{{Lemson} \& {Kauffmann}}{{Lemson} \&
  {Kauffmann}}{1999}]{Lemson_1999}
{Lemson} G.,  {Kauffmann} G.,  1999, \mn@doi [\mnras]
  {10.1046/j.1365-8711.1999.02090.x}, \href
  {http://adsabs.harvard.edu/abs/1999MNRAS.302..111L} {302, 111}

\bibitem[\protect\citeauthoryear{{Lewis} \& {Challinor}}{{Lewis} \&
  {Challinor}}{2011}]{Lewis_2011}
{Lewis} A.,  {Challinor} A.,  2011, {CAMB: Code for Anisotropies in the
  Microwave Background}, Astrophysics Source Code Library (\mn@eprint {ascl}
  {1102.026})

\bibitem[\protect\citeauthoryear{{Mandelbaum} et~al.,}{{Mandelbaum}
  et~al.}{2005}]{Mandelbaum_2005}
{Mandelbaum} R.,  et~al., 2005, \mn@doi [\mnras]
  {10.1111/j.1365-2966.2005.09282.x}, \href
  {http://adsabs.harvard.edu/abs/2005MNRAS.361.1287M} {361, 1287}

\bibitem[\protect\citeauthoryear{{Mandelbaum}, {Seljak}, {Cool}, {Blanton},
  {Hirata}  \& {Brinkmann}}{{Mandelbaum} et~al.}{2006}]{Mandelbaum_2006}
{Mandelbaum} R.,  {Seljak} U.,  {Cool} R.~J.,  {Blanton} M.,  {Hirata} C.~M.,
  {Brinkmann} J.,  2006, \mn@doi [\mnras] {10.1111/j.1365-2966.2006.10906.x},
  \href {http://adsabs.harvard.edu/abs/2006MNRAS.372..758M} {372, 758}

\bibitem[\protect\citeauthoryear{{Mandelbaum}, {Slosar}, {Baldauf}, {Seljak},
  {Hirata}, {Nakajima}, {Reyes}  \& {Smith}}{{Mandelbaum}
  et~al.}{2013}]{Mandelbaum_2013}
{Mandelbaum} R.,  {Slosar} A.,  {Baldauf} T.,  {Seljak} U.,  {Hirata} C.~M.,
  {Nakajima} R.,  {Reyes} R.,   {Smith} R.~E.,  2013, \mn@doi [\mnras]
  {10.1093/mnras/stt572}, \href
  {http://adsabs.harvard.edu/abs/2013MNRAS.432.1544M} {432, 1544}

\bibitem[\protect\citeauthoryear{{McEwen} \& {Weinberg}}{{McEwen} \&
  {Weinberg}}{2016}]{McEwen_2016}
{McEwen} J.~E.,  {Weinberg} D.~H.,  2016, preprint, \href
  {http://adsabs.harvard.edu/abs/2016arXiv160102693M} {} (\mn@eprint {arXiv}
  {1601.02693})

\bibitem[\protect\citeauthoryear{{Metchnik}}{{Metchnik}}{2009}]{Metchnik_2009}
{Metchnik} M.~V.~L.,  2009, PhD thesis, The University of Arizona

\bibitem[\protect\citeauthoryear{{More}}{{More}}{2013}]{More_2013b}
{More} S.,  2013, \mn@doi [\apjl] {10.1088/2041-8205/777/2/L26}, \href
  {http://adsabs.harvard.edu/abs/2013ApJ...777L..26M} {777, L26}

\bibitem[\protect\citeauthoryear{{More}, {van den Bosch}, {Cacciato}, {More},
  {Mo}  \& {Yang}}{{More} et~al.}{2013}]{More_2013a}
{More} S.,  {van den Bosch} F.~C.,  {Cacciato} M.,  {More} A.,  {Mo} H.,
  {Yang} X.,  2013, \mn@doi [\mnras] {10.1093/mnras/sts697}, \href
  {http://adsabs.harvard.edu/abs/2013MNRAS.430..747M} {430, 747}

\bibitem[\protect\citeauthoryear{{More}, {Miyatake}, {Mandelbaum}, {Takada},
  {Spergel}, {Brownstein}  \& {Schneider}}{{More} et~al.}{2015}]{More_2015}
{More} S.,  {Miyatake} H.,  {Mandelbaum} R.,  {Takada} M.,  {Spergel} D.~N.,
  {Brownstein} J.~R.,   {Schneider} D.~P.,  2015, \mn@doi [\apj]
  {10.1088/0004-637X/806/1/2}, \href
  {http://adsabs.harvard.edu/abs/2015ApJ...806....2M} {806, 2}

\bibitem[\protect\citeauthoryear{{Navarro}, {Frenk}  \& {White}}{{Navarro}
  et~al.}{1997}]{NFW_1997}
{Navarro} J.~F.,  {Frenk} C.~S.,   {White} S.~D.~M.,  1997, \mn@doi [\apj]
  {10.1086/304888}, \href {http://adsabs.harvard.edu/abs/1997ApJ...490..493N}
  {490, 493}

\bibitem[\protect\citeauthoryear{{Parejko} et~al.,}{{Parejko}
  et~al.}{2013}]{Parejko_2013}
{Parejko} J.~K.,  et~al., 2013, \mn@doi [\mnras] {10.1093/mnras/sts314}, \href
  {http://adsabs.harvard.edu/abs/2013MNRAS.429...98P} {429, 98}

\bibitem[\protect\citeauthoryear{{Planck Collaboration} et~al.,}{{Planck
  Collaboration} et~al.}{2016}]{Planck_2016}
{Planck Collaboration} et~al., 2016, \mn@doi [\aap]
  {10.1051/0004-6361/201525830}, \href
  {http://adsabs.harvard.edu/abs/2016A%26A...594A..13P} {594, A13}

\bibitem[\protect\citeauthoryear{{Prat} et~al.,}{{Prat}
  et~al.}{2017}]{Prat_2017}
{Prat} J.,  et~al., 2017, preprint, \href
  {http://adsabs.harvard.edu/abs/2017arXiv170801537P} {} (\mn@eprint {arXiv}
  {1708.01537})

\bibitem[\protect\citeauthoryear{{Rozo} et~al.,}{{Rozo}
  et~al.}{2016}]{Rozo_2016}
{Rozo} E.,  et~al., 2016, \mn@doi [\mnras] {10.1093/mnras/stw1281}, \href
  {http://adsabs.harvard.edu/abs/2016MNRAS.461.1431R} {461, 1431}

\bibitem[\protect\citeauthoryear{{Salcedo}, {Maller}, {Berlind}, {Sinha},
  {McBride}, {Behroozi}, {Wechsler}  \& {Weinberg}}{{Salcedo}
  et~al.}{2017}]{Salcedo_2017}
{Salcedo} A.~N.,  {Maller} A.~H.,  {Berlind} A.~A.,  {Sinha} M.,  {McBride}
  C.~K.,  {Behroozi} P.~S.,  {Wechsler} R.~H.,   {Weinberg} D.~H.,  2017,
  preprint, \href {http://adsabs.harvard.edu/abs/2017arXiv170808451S} {}
  (\mn@eprint {arXiv} {1708.08451})

\bibitem[\protect\citeauthoryear{{Scoccimarro}, {Zaldarriaga}  \&
  {Hui}}{{Scoccimarro} et~al.}{1999}]{Scoccimarro_1999}
{Scoccimarro} R.,  {Zaldarriaga} M.,   {Hui} L.,  1999, \mn@doi [\apj]
  {10.1086/308059}, \href {http://adsabs.harvard.edu/abs/1999ApJ...527....1S}
  {527, 1}

\bibitem[\protect\citeauthoryear{{Sheldon} et~al.,}{{Sheldon}
  et~al.}{2004}]{Sheldon_2004}
{Sheldon} E.~S.,  et~al., 2004, \mn@doi [\aj] {10.1086/383293}, \href
  {http://adsabs.harvard.edu/abs/2004AJ....127.2544S} {127, 2544}

\bibitem[\protect\citeauthoryear{{Sheth} \& {Tormen}}{{Sheth} \&
  {Tormen}}{2004}]{Sheth_2004}
{Sheth} R.~K.,  {Tormen} G.,  2004, \mn@doi [\mnras]
  {10.1111/j.1365-2966.2004.07733.x}, \href
  {http://adsabs.harvard.edu/abs/2004MNRAS.350.1385S} {350, 1385}

\bibitem[\protect\citeauthoryear{{Singh}, {Mandelbaum}, {Seljak}, {Slosar}  \&
  {Vazquez Gonzalez}}{{Singh} et~al.}{2016}]{Singh_2016}
{Singh} S.,  {Mandelbaum} R.,  {Seljak} U.,  {Slosar} A.,   {Vazquez Gonzalez}
  J.,  2016, preprint, \href
  {http://adsabs.harvard.edu/abs/2016arXiv161100752S} {} (\mn@eprint {arXiv}
  {1611.00752})

\bibitem[\protect\citeauthoryear{{Sinha} \& {Garrison}}{{Sinha} \&
  {Garrison}}{2017}]{Sinha_2017}
{Sinha} M.,  {Garrison} L.,  2017, {Corrfunc: Blazing fast correlation
  functions on the CPU}, Astrophysics Source Code Library (\mn@eprint {ascl}
  {1703.003})

\bibitem[\protect\citeauthoryear{{Tojeiro} et~al.,}{{Tojeiro}
  et~al.}{2014}]{Tojeiro_2014}
{Tojeiro} R.,  et~al., 2014, \mn@doi [\mnras] {10.1093/mnras/stu371}, \href
  {http://adsabs.harvard.edu/abs/2014MNRAS.440.2222T} {440, 2222}

\bibitem[\protect\citeauthoryear{{Troxel} et~al.,}{{Troxel}
  et~al.}{2017}]{Troxel_2017}
{Troxel} M.~A.,  et~al., 2017, preprint, \href
  {http://adsabs.harvard.edu/abs/2017arXiv170801538T} {} (\mn@eprint {arXiv}
  {1708.01538})

\bibitem[\protect\citeauthoryear{{Wechsler}, {Zentner}, {Bullock}, {Kravtsov}
  \& {Allgood}}{{Wechsler} et~al.}{2006}]{Wechsler_2006}
{Wechsler} R.~H.,  {Zentner} A.~R.,  {Bullock} J.~S.,  {Kravtsov} A.~V.,
  {Allgood} B.,  2006, \mn@doi [\apj] {10.1086/507120}, \href
  {http://adsabs.harvard.edu/abs/2006ApJ...652...71W} {652, 71}

\bibitem[\protect\citeauthoryear{{Weinberg}, {Mortonson}, {Eisenstein},
  {Hirata}, {Riess}  \& {Rozo}}{{Weinberg} et~al.}{2013}]{Weinberg_2013}
{Weinberg} D.~H.,  {Mortonson} M.~J.,  {Eisenstein} D.~J.,  {Hirata} C.,
  {Riess} A.~G.,   {Rozo} E.,  2013, \mn@doi [\physrep]
  {10.1016/j.physrep.2013.05.001}, \href
  {http://adsabs.harvard.edu/abs/2013PhR...530...87W} {530, 87}

\bibitem[\protect\citeauthoryear{{White}}{{White}}{1994}]{White_1994}
{White} S.~D.~M.,  1994, ArXiv Astrophysics e-prints, \href
  {http://adsabs.harvard.edu/abs/1994astro.ph.10043W} {}

\bibitem[\protect\citeauthoryear{{Yoo} \& {Seljak}}{{Yoo} \&
  {Seljak}}{2012}]{Yoo_2012}
{Yoo} J.,  {Seljak} U.,  2012, \mn@doi [\prd] {10.1103/PhysRevD.86.083504},
  \href {http://adsabs.harvard.edu/abs/2012PhRvD..86h3504Y} {86, 083504}

\bibitem[\protect\citeauthoryear{{Yoo}, {Tinker}, {Weinberg}, {Zheng}, {Katz}
  \& {Dav{\'e}}}{{Yoo} et~al.}{2006}]{Yoo_2006}
{Yoo} J.,  {Tinker} J.~L.,  {Weinberg} D.~H.,  {Zheng} Z.,  {Katz} N.,
  {Dav{\'e}} R.,  2006, \mn@doi [\apj] {10.1086/507591}, \href
  {http://adsabs.harvard.edu/abs/2006ApJ...652...26Y} {652, 26}

\bibitem[\protect\citeauthoryear{{York} et~al.,}{{York}
  et~al.}{2000}]{York_2000}
{York} D.~G.,  et~al., 2000, \mn@doi [\aj] {10.1086/301513}, \href
  {http://adsabs.harvard.edu/abs/2000AJ....120.1579Y} {120, 1579}

\bibitem[\protect\citeauthoryear{{Zehavi} et~al.,}{{Zehavi}
  et~al.}{2005}]{Zehavi_2005}
{Zehavi} I.,  et~al., 2005, \mn@doi [\apj] {10.1086/431891}, \href
  {http://adsabs.harvard.edu/abs/2005ApJ...630....1Z} {630, 1}

\bibitem[\protect\citeauthoryear{{Zehavi} et~al.,}{{Zehavi}
  et~al.}{2011}]{Zehavi_2011}
{Zehavi} I.,  et~al., 2011, \mn@doi [\apj] {10.1088/0004-637X/736/1/59}, \href
  {http://adsabs.harvard.edu/abs/2011ApJ...736...59Z} {736, 59}

\bibitem[\protect\citeauthoryear{{Zentner}, {Hearin}  \& {van den
  Bosch}}{{Zentner} et~al.}{2014}]{Zentner_2014}
{Zentner} A.~R.,  {Hearin} A.~P.,   {van den Bosch} F.~C.,  2014, \mn@doi
  [\mnras] {10.1093/mnras/stu1383}, \href
  {http://adsabs.harvard.edu/abs/2014MNRAS.443.3044Z} {443, 3044}

\bibitem[\protect\citeauthoryear{{Zheng} et~al.,}{{Zheng}
  et~al.}{2005}]{Zheng_2005}
{Zheng} Z.,  et~al., 2005, \mn@doi [\apj] {10.1086/466510}, \href
  {http://adsabs.harvard.edu/abs/2005ApJ...633..791Z} {633, 791}

\bibitem[\protect\citeauthoryear{{Zu} \& {Mandelbaum}}{{Zu} \&
  {Mandelbaum}}{2015}]{Zu_2015}
{Zu} Y.,  {Mandelbaum} R.,  2015, \mn@doi [\mnras] {10.1093/mnras/stv2062},
  \href {http://adsabs.harvard.edu/abs/2015MNRAS.454.1161Z} {454, 1161}

\bibitem[\protect\citeauthoryear{{van den Bosch}, {More}, {Cacciato}, {Mo}  \&
  {Yang}}{{van den Bosch} et~al.}{2013}]{vdBosch_2013}
{van den Bosch} F.~C.,  {More} S.,  {Cacciato} M.,  {Mo} H.,   {Yang} X.,
  2013, \mn@doi [\mnras] {10.1093/mnras/sts006}, \href
  {http://adsabs.harvard.edu/abs/2013MNRAS.430..725V} {430, 725}

\makeatother
\end{thebibliography}



\onecolumn
\appendix
\section{Emulator Derivatives}
\label{appendix:derivatives}

\begin{longtabu} to \textwidth {*{11}{X[1,r]}}
\caption{Logarithm of the galaxy bias $\ln b_{\text{g}}(r)$ and its partial derivatives with respect to HOD and cosmological parameters, tabulated in radial bins of width $\Delta r_i = r_{i+1} - r_i$. (The radial separation $r$ is in units of $h^{-1}$ Mpc.)} \\
\toprule
$r_i$ & $r_{i+1}$ & fiducial & $\frac{\partial}{\partial \ln n_{\text{gal}}}$ & $\frac{\partial}{\partial \ln \sigma_{\log M}}$ & $\frac{\partial}{\partial \ln M_1 / M_{\text{min}}}$ & $\frac{\partial}{\partial \ln \alpha}$ & $\frac{\partial}{\partial Q_{\text{env}}}$ & $\frac{\partial}{\partial \Delta \gamma}$ & $\frac{\partial}{\partial \ln \Omega_M}$ & $\frac{\partial}{\partial \ln \sigma_8}$ \\
\midrule
\endfirsthead
\toprule
$r_i$ & $r_{i+1}$ & fiducial & $\frac{\partial}{\partial \ln n_{\text{gal}}}$ & $\frac{\partial}{\partial \ln \sigma_{\log M}}$ & $\frac{\partial}{\partial \ln M_1 / M_{\text{min}}}$ & $\frac{\partial}{\partial \ln \alpha}$ & $\frac{\partial}{\partial Q_{\text{env}}}$ & $\frac{\partial}{\partial \Delta \gamma}$ & $\frac{\partial}{\partial \ln \Omega_M}$ & $\frac{\partial}{\partial \ln \sigma_8}$ \\
\midrule
\endhead
0.0100 & 0.0108 & 2.2939 & -1.4216 & -1.5205 & -0.0136 & -0.4796 & 0.4149 & -2.0304 & 1.0275 & -4.3543 \\
0.0108 & 0.0117 & 2.3555 & -0.9990 & -1.0520 & -0.2934 & -1.7703 & 0.2218 & -1.0442 & -2.7467 & 0.6173 \\
0.0117 & 0.0127 & 2.0129 & -0.9995 & -1.2740 & -1.3748 & -0.8794 & -0.0722 & -1.8599 & 1.1547 & -1.8179 \\
0.0127 & 0.0137 & 2.6067 & 1.0120 & -1.0876 & -0.4934 & -0.2416 & -1.0150 & -1.8536 & 0.5864 & 0.4710 \\
0.0137 & 0.0148 & 2.1467 & 0.0533 & -0.5756 & -2.1999 & -0.8930 & 0.3733 & -1.6124 & -0.6663 & -0.5664 \\
0.0148 & 0.0160 & 2.0787 & -0.2391 & -0.5046 & -1.5715 & -0.4700 & 0.5684 & -1.5997 & 1.2111 & 0.0846 \\
0.0160 & 0.0173 & 2.0617 & 0.3168 & -1.9444 & -1.7972 & -0.5000 & -0.1176 & -1.2653 & -0.2406 & 0.3321 \\
0.0173 & 0.0188 & 1.9884 & -0.2533 & -1.5436 & -0.8169 & -0.8556 & 0.0337 & -1.1229 & -1.9151 & -0.8290 \\
0.0188 & 0.0203 & 2.0849 & 0.2785 & -1.3071 & -0.2104 & -0.6600 & -0.3194 & -2.0445 & -0.4284 & -0.7180 \\
0.0203 & 0.0219 & 2.1242 & 0.1452 & -1.3336 & -0.0006 & -0.7713 & 0.2923 & -0.8610 & 0.7507 & -1.3470 \\
0.0219 & 0.0237 & 1.9130 & -0.2992 & -1.2411 & -1.3735 & -0.8943 & 0.0029 & -1.5227 & 0.3867 & -1.2434 \\
0.0237 & 0.0257 & 1.9217 & -0.0511 & -1.1796 & -1.4125 & -0.9768 & 0.2523 & -1.5630 & 0.2135 & -0.9692 \\
0.0257 & 0.0278 & 1.8913 & 0.1532 & -1.2036 & -1.3560 & -0.6439 & 0.4940 & -1.4629 & -0.3560 & -1.0613 \\
0.0278 & 0.0301 & 1.8178 & -0.6734 & -1.2754 & -0.8683 & -0.9012 & -0.1444 & -1.1220 & -0.1319 & -2.4338 \\
0.0301 & 0.0325 & 1.6636 & -0.4060 & -1.2594 & -1.2049 & -0.7541 & -0.0893 & -1.2352 & -0.4601 & -1.7240 \\
0.0325 & 0.0352 & 1.7146 & -0.2379 & -1.1582 & -1.4256 & -0.4115 & 0.0689 & -1.1822 & -0.5567 & -0.6219 \\
0.0352 & 0.0381 & 1.6478 & -0.1291 & -1.1537 & -0.7704 & -0.8863 & -0.1367 & -1.4389 & 0.1470 & -0.8603 \\
0.0381 & 0.0412 & 1.5927 & 0.4057 & -1.0964 & -0.7063 & -0.6317 & 0.2485 & -1.0942 & -0.0123 & -1.1123 \\
0.0412 & 0.0445 & 1.5502 & -0.0875 & -1.1355 & -1.5382 & -0.6579 & 0.0272 & -1.0979 & 0.1520 & -0.4945 \\
0.0445 & 0.0482 & 1.4942 & 0.0105 & -1.2977 & -0.6850 & -0.8993 & -0.3457 & -1.1600 & 0.2238 & -0.8423 \\
0.0482 & 0.0521 & 1.4409 & 0.0347 & -1.1579 & -0.8967 & -0.7894 & -0.3028 & -0.9494 & 0.3722 & -0.4377 \\
0.0521 & 0.0564 & 1.4456 & -0.0208 & -1.3066 & -1.2969 & -0.8550 & 0.3199 & -1.1406 & -0.4201 & -1.0993 \\
0.0564 & 0.0610 & 1.4100 & -0.1737 & -1.1600 & -0.3600 & -0.6369 & -0.0393 & -1.0696 & -0.3380 & -1.3311 \\
0.0610 & 0.0660 & 1.3958 & 0.1130 & -1.2796 & -0.6179 & -0.5716 & 0.1474 & -0.8697 & 0.0728 & -0.6497 \\
0.0660 & 0.0714 & 1.3069 & -0.0592 & -1.1141 & -1.1611 & -0.7315 & -0.1852 & -0.9905 & -0.1474 & -0.8880 \\
0.0714 & 0.0772 & 1.3152 & -0.0802 & -1.3554 & -0.9373 & -0.6703 & 0.0366 & -0.9441 & -0.1790 & -0.6032 \\
0.0772 & 0.0835 & 1.2744 & -0.0273 & -1.2151 & -0.8925 & -0.8273 & -0.1377 & -0.9422 & -0.0913 & -0.7371 \\
0.0835 & 0.0904 & 1.2433 & -0.1332 & -1.2738 & -0.6406 & -0.6925 & 0.0915 & -0.5834 & -0.0359 & -0.9176 \\
0.0904 & 0.0977 & 1.1666 & -0.2304 & -1.4296 & -0.8622 & -0.5022 & -0.1516 & -0.5751 & 0.0227 & -1.1295 \\
0.0977 & 0.1057 & 1.1161 & -0.0672 & -1.2836 & -0.7687 & -0.4688 & 0.0224 & -0.7863 & 0.0236 & -0.6780 \\
0.1057 & 0.1144 & 1.0596 & -0.0665 & -1.3129 & -0.7337 & -0.4890 & 0.0156 & -0.7300 & 0.1208 & -0.4225 \\
0.1144 & 0.1237 & 1.0777 & -0.1417 & -1.2135 & -0.6592 & -0.7034 & -0.0172 & -0.6171 & -0.0100 & -0.7293 \\
0.1237 & 0.1339 & 1.0061 & -0.0365 & -1.0818 & -0.9693 & -0.5437 & -0.1599 & -0.6153 & -0.0843 & -0.6931 \\
0.1339 & 0.1448 & 0.9826 & 0.0048 & -1.1168 & -0.4576 & -0.4726 & 0.0821 & -0.5270 & 0.0765 & -0.8382 \\
0.1448 & 0.1567 & 0.9516 & -0.1569 & -1.1390 & -1.0246 & -0.5796 & -0.0487 & -0.5330 & -0.0755 & -0.9601 \\
0.1567 & 0.1695 & 0.9069 & -0.0748 & -1.1262 & -0.9459 & -0.5865 & -0.0200 & -0.4502 & -0.1288 & -0.5235 \\
0.1695 & 0.1833 & 0.8695 & -0.0414 & -1.1608 & -0.6565 & -0.5488 & 0.0057 & -0.5486 & 0.0162 & -0.8356 \\
0.1833 & 0.1983 & 0.8457 & -0.2059 & -1.1984 & -0.6828 & -0.5913 & 0.0661 & -0.4660 & -0.0701 & -0.8375 \\
0.1983 & 0.2145 & 0.8180 & -0.1828 & -1.1210 & -0.8152 & -0.5545 & 0.0003 & -0.3977 & 0.1479 & -0.5073 \\
0.2145 & 0.2321 & 0.7786 & 0.0133 & -1.1437 & -0.9786 & -0.5578 & 0.0228 & -0.2677 & 0.0460 & -0.6997 \\
0.2321 & 0.2511 & 0.7400 & -0.1869 & -1.2059 & -0.8878 & -0.5653 & 0.0188 & -0.2367 & 0.0588 & -0.4721 \\
0.2511 & 0.2716 & 0.6942 & -0.0833 & -1.1775 & -0.8700 & -0.5439 & 0.0994 & -0.3736 & -0.1053 & -0.6909 \\
0.2716 & 0.2938 & 0.6896 & -0.1488 & -1.1255 & -0.8372 & -0.5569 & 0.0630 & -0.2821 & -0.0707 & -0.6425 \\
0.2938 & 0.3178 & 0.6494 & -0.0738 & -1.1102 & -0.8108 & -0.5252 & 0.0585 & -0.1731 & 0.0116 & -0.6716 \\
0.3178 & 0.3438 & 0.6076 & -0.0960 & -1.1076 & -0.9646 & -0.4541 & 0.0052 & -0.2124 & -0.0568 & -0.5808 \\
0.3438 & 0.3719 & 0.6014 & -0.0148 & -1.1216 & -0.5694 & -0.5036 & -0.0835 & -0.1993 & -0.0765 & -0.5565 \\
0.3719 & 0.4024 & 0.5665 & -0.1632 & -1.1236 & -0.9424 & -0.4555 & 0.1165 & -0.0561 & -0.0669 & -0.7945 \\
0.4024 & 0.4353 & 0.5425 & -0.0477 & -1.0826 & -0.8063 & -0.3838 & 0.0038 & -0.1187 & -0.1290 & -0.7698 \\
0.4353 & 0.4709 & 0.5416 & 0.0142 & -1.0247 & -0.6920 & -0.4427 & 0.0185 & -0.0246 & -0.0961 & -0.7389 \\
0.4709 & 0.5094 & 0.5195 & -0.0737 & -1.0333 & -0.7976 & -0.3779 & -0.0127 & -0.0326 & -0.2265 & -0.7183 \\
0.5094 & 0.5510 & 0.5188 & 0.0156 & -0.9331 & -0.5879 & -0.3615 & -0.0527 & 0.0796 & -0.1142 & -0.8013 \\
0.5510 & 0.5961 & 0.4993 & 0.0334 & -0.9327 & -0.7192 & -0.3347 & -0.0640 & 0.0870 & -0.0860 & -0.9196 \\
0.5961 & 0.6449 & 0.5088 & -0.0596 & -0.8841 & -0.6131 & -0.2954 & -0.0018 & 0.0704 & -0.1330 & -1.0319 \\
0.6449 & 0.6976 & 0.5146 & -0.0785 & -0.8813 & -0.6356 & -0.2570 & -0.0958 & 0.0363 & -0.1195 & -1.1053 \\
0.6976 & 0.7547 & 0.5476 & -0.0518 & -0.8121 & -0.5102 & -0.2525 & -0.1252 & 0.0924 & -0.1102 & -1.0091 \\
0.7547 & 0.8164 & 0.5445 & -0.0981 & -0.7618 & -0.5256 & -0.1536 & -0.1293 & 0.0858 & -0.0776 & -1.2019 \\
0.8164 & 0.8831 & 0.5561 & -0.1296 & -0.7186 & -0.4822 & -0.1257 & -0.1912 & 0.1713 & -0.1051 & -1.0256 \\
0.8831 & 0.9554 & 0.5485 & -0.1495 & -0.6748 & -0.4344 & -0.1022 & -0.1912 & 0.1064 & -0.1698 & -1.1144 \\
0.9554 & 1.0335 & 0.5463 & -0.1599 & -0.6705 & -0.3657 & -0.0408 & -0.2211 & 0.1160 & -0.1304 & -0.9979 \\
1.0335 & 1.1180 & 0.5483 & -0.1806 & -0.6061 & -0.3762 & 0.0115 & -0.2550 & 0.1135 & -0.1447 & -0.9510 \\
1.1180 & 1.2095 & 0.5334 & -0.1737 & -0.5753 & -0.3881 & 0.0235 & -0.2619 & 0.1166 & -0.1194 & -1.0430 \\
1.2095 & 1.3084 & 0.5396 & -0.1732 & -0.5047 & -0.3223 & 0.0405 & -0.2965 & 0.0890 & -0.1782 & -1.0498 \\
1.3084 & 1.4154 & 0.5255 & -0.1557 & -0.4847 & -0.2932 & 0.1031 & -0.3182 & 0.0564 & -0.1235 & -1.0985 \\
1.4154 & 1.5312 & 0.5246 & -0.1219 & -0.4800 & -0.2177 & 0.1071 & -0.3524 & 0.0852 & -0.0415 & -1.0936 \\
1.5312 & 1.6564 & 0.5270 & -0.1508 & -0.4296 & -0.2502 & 0.0728 & -0.3716 & 0.0615 & -0.0641 & -1.0821 \\
1.6564 & 1.7918 & 0.5254 & -0.1204 & -0.4213 & -0.2319 & 0.0833 & -0.4424 & 0.0266 & -0.1118 & -1.2226 \\
1.7918 & 1.9384 & 0.5290 & -0.1594 & -0.3766 & -0.1973 & 0.0760 & -0.4247 & 0.0467 & -0.0776 & -1.1909 \\
1.9384 & 2.0969 & 0.5600 & -0.1732 & -0.3851 & -0.1642 & 0.0840 & -0.4436 & 0.0239 & -0.1279 & -1.2412 \\
2.0969 & 2.2684 & 0.5597 & -0.1418 & -0.3636 & -0.1391 & 0.0520 & -0.4488 & 0.0157 & -0.0013 & -1.1663 \\
2.2684 & 2.4539 & 0.5514 & -0.1195 & -0.3384 & -0.0918 & 0.0508 & -0.5090 & 0.0280 & -0.0603 & -1.2421 \\
2.4539 & 2.6546 & 0.5743 & -0.1404 & -0.3188 & -0.1169 & 0.0200 & -0.4932 & 0.0040 & 0.0060 & -0.9666 \\
2.6546 & 2.8717 & 0.5802 & -0.1363 & -0.3258 & -0.1018 & 0.0263 & -0.5165 & 0.0141 & -0.0077 & -1.0896 \\
2.8717 & 3.1066 & 0.5852 & -0.1450 & -0.3119 & -0.0970 & -0.0041 & -0.5244 & 0.0078 & 0.0477 & -1.0441 \\
3.1066 & 3.3607 & 0.5852 & -0.1234 & -0.3001 & -0.0899 & 0.0105 & -0.5438 & 0.0029 & -0.1317 & -0.8835 \\
3.3607 & 3.6355 & 0.5883 & -0.1307 & -0.3238 & -0.0771 & -0.0104 & -0.5592 & 0.0036 & -0.0435 & -0.8820 \\
3.6355 & 3.9329 & 0.5942 & -0.1402 & -0.3069 & -0.0757 & -0.0033 & -0.5732 & 0.0049 & -0.0066 & -0.8703 \\
3.9329 & 4.2545 & 0.5948 & -0.1290 & -0.2945 & -0.0777 & -0.0075 & -0.6119 & -0.0013 & -0.0175 & -0.9219 \\
4.2545 & 4.6025 & 0.5936 & -0.1445 & -0.3078 & -0.0677 & -0.0082 & -0.6046 & 0.0014 & -0.0700 & -0.8414 \\
4.6025 & 4.9789 & 0.5876 & -0.1394 & -0.3014 & -0.0865 & -0.0092 & -0.6399 & 0.0022 & -0.0503 & -0.7890 \\
4.9789 & 5.3861 & 0.5847 & -0.1412 & -0.3067 & -0.1029 & -0.0156 & -0.6678 & 0.0006 & -0.0198 & -0.8344 \\
5.3861 & 5.8266 & 0.5759 & -0.1436 & -0.2989 & -0.0934 & -0.0110 & -0.7140 & -0.0006 & -0.0242 & -0.7452 \\
5.8266 & 6.3031 & 0.5779 & -0.1339 & -0.3020 & -0.0878 & -0.0176 & -0.7375 & -0.0018 & -0.0234 & -0.7402 \\
6.3031 & 6.8186 & 0.5843 & -0.1466 & -0.3038 & -0.0843 & -0.0090 & -0.7409 & 0.0012 & -0.0175 & -0.7003 \\
6.8186 & 7.3763 & 0.5776 & -0.1390 & -0.3029 & -0.1029 & -0.0125 & -0.7775 & -0.0002 & -0.0714 & -0.7106 \\
7.3763 & 7.9795 & 0.5757 & -0.1473 & -0.3030 & -0.0778 & -0.0060 & -0.7586 & 0.0008 & -0.0248 & -0.7507 \\
7.9795 & 8.6321 & 0.5670 & -0.1393 & -0.2895 & -0.0633 & -0.0042 & -0.6561 & -0.0014 & -0.0037 & -0.7310 \\
8.6321 & 9.3381 & 0.5589 & -0.1437 & -0.2908 & -0.0569 & -0.0139 & -0.6002 & 0.0019 & -0.0285 & -0.8025 \\
9.3381 & 10.1018 & 0.5567 & -0.1446 & -0.2903 & -0.0545 & -0.0108 & -0.5900 & 0.0004 & 0.0127 & -0.6848 \\
10.1018 & 10.9280 & 0.5494 & -0.1343 & -0.2802 & -0.0663 & -0.0130 & -0.5702 & -0.0007 & 0.0028 & -0.8109 \\
10.9280 & 11.8218 & 0.5460 & -0.1460 & -0.2734 & -0.0593 & -0.0082 & -0.5611 & 0.0006 & -0.0166 & -0.7310 \\
11.8218 & 12.7886 & 0.5454 & -0.1317 & -0.2777 & -0.0896 & -0.0180 & -0.5627 & -0.0005 & -0.0363 & -0.7387 \\
12.7886 & 13.8345 & 0.5442 & -0.1320 & -0.2886 & -0.0754 & -0.0151 & -0.5490 & -0.0001 & -0.0271 & -0.7487 \\
13.8345 & 14.9660 & 0.5463 & -0.1426 & -0.2875 & -0.0848 & -0.0118 & -0.5384 & -0.0002 & -0.0110 & -0.7359 \\
14.9660 & 16.1900 & 0.5479 & -0.1330 & -0.2852 & -0.0864 & -0.0081 & -0.5182 & 0.0006 & 0.0097 & -0.7896 \\
16.1900 & 17.5141 & 0.5410 & -0.1284 & -0.2795 & -0.0833 & -0.0047 & -0.5194 & 0.0005 & -0.0183 & -0.7284 \\
17.5141 & 18.9465 & 0.5468 & -0.1358 & -0.2816 & -0.0691 & -0.0122 & -0.5335 & -0.0002 & -0.0093 & -0.7329 \\
18.9465 & 20.4960 & 0.5468 & -0.1293 & -0.2847 & -0.0647 & -0.0155 & -0.5243 & -0.0007 & -0.0007 & -0.7629 \\
20.4960 & 22.1723 & 0.5448 & -0.1363 & -0.2889 & -0.0651 & -0.0155 & -0.5258 & 0.0002 & -0.0304 & -0.7878 \\
22.1723 & 23.9856 & 0.5332 & -0.1154 & -0.2870 & -0.0592 & -0.0036 & -0.5557 & 0.0002 & -0.0654 & -0.7387 \\
23.9856 & 25.9473 & 0.5361 & -0.1211 & -0.2889 & -0.0614 & -0.0026 & -0.5426 & 0.0003 & -0.0136 & -0.6789 \\
25.9473 & 28.0694 & 0.5462 & -0.1329 & -0.2942 & -0.0473 & -0.0120 & -0.5552 & -0.0008 & -0.0466 & -0.6378 \\
28.0694 & 30.3650 & 0.5460 & -0.1339 & -0.2852 & -0.0682 & -0.0110 & -0.5499 & 0.0001 & -0.0185 & -0.7408 \\
30.3650 & 32.8484 & 0.5460 & -0.1339 & -0.2852 & -0.0682 & -0.0110 & -0.5499 & 0.0001 & -0.0185 & -0.7408 \\
32.8484 & 35.5349 & 0.5460 & -0.1339 & -0.2852 & -0.0682 & -0.0110 & -0.5499 & 0.0001 & -0.0185 & -0.7408 \\
35.5349 & 38.4411 & 0.5460 & -0.1339 & -0.2852 & -0.0682 & -0.0110 & -0.5499 & 0.0001 & -0.0185 & -0.7408 \\
38.4411 & 41.5850 & 0.5460 & -0.1339 & -0.2852 & -0.0682 & -0.0110 & -0.5499 & 0.0001 & -0.0185 & -0.7408 \\
41.5850 & 44.9861 & 0.5460 & -0.1339 & -0.2852 & -0.0682 & -0.0110 & -0.5499 & 0.0001 & -0.0185 & -0.7408 \\
44.9861 & 48.6653 & 0.5460 & -0.1339 & -0.2852 & -0.0682 & -0.0110 & -0.5499 & 0.0001 & -0.0185 & -0.7408 \\
48.6653 & 52.6453 & 0.5460 & -0.1339 & -0.2852 & -0.0682 & -0.0110 & -0.5499 & 0.0001 & -0.0185 & -0.7408 \\
52.6453 & 56.9509 & 0.5460 & -0.1339 & -0.2852 & -0.0682 & -0.0110 & -0.5499 & 0.0001 & -0.0185 & -0.7408 \\
56.9509 & 61.6087 & 0.5460 & -0.1339 & -0.2852 & -0.0682 & -0.0110 & -0.5499 & 0.0001 & -0.0185 & -0.7408 \\
61.6087 & 66.6473 & 0.5460 & -0.1339 & -0.2852 & -0.0682 & -0.0110 & -0.5499 & 0.0001 & -0.0185 & -0.7408 \\
66.6473 & 72.0981 & 0.5460 & -0.1339 & -0.2852 & -0.0682 & -0.0110 & -0.5499 & 0.0001 & -0.0185 & -0.7408 \\
72.0981 & 77.9946 & 0.5460 & -0.1339 & -0.2852 & -0.0682 & -0.0110 & -0.5499 & 0.0001 & -0.0185 & -0.7408 \\
77.9946 & 84.3734 & 0.5460 & -0.1339 & -0.2852 & -0.0682 & -0.0110 & -0.5499 & 0.0001 & -0.0185 & -0.7408 \\
84.3734 & 91.2738 & 0.5460 & -0.1339 & -0.2852 & -0.0682 & -0.0110 & -0.5499 & 0.0001 & -0.0185 & -0.7408 \\
91.2738 & 98.7387 & 0.5460 & -0.1339 & -0.2852 & -0.0682 & -0.0110 & -0.5499 & 0.0001 & -0.0185 & -0.7408 \\
98.7387 & 106.8140 & 0.5460 & -0.1339 & -0.2852 & -0.0682 & -0.0110 & -0.5499 & 0.0001 & -0.0185 & -0.7408 \\
106.8140 & 115.5498 & 0.5460 & -0.1339 & -0.2852 & -0.0682 & -0.0110 & -0.5499 & 0.0001 & -0.0185 & -0.7408 \\
115.5498 & 125.0000 & 0.5460 & -0.1339 & -0.2852 & -0.0682 & -0.0110 & -0.5499 & 0.0001 & -0.0185 & -0.7408 \\
\bottomrule
\end{longtabu}

\begin{longtabu} to \textwidth {*{11}{X[1,r]}}
\caption{Logarithm of the galaxy-matter correlation coefficient $\ln r_{\text{gm}}(r)$ and its partial derivatives with respect to cosmological parameters, tabulated in radial bins of width $\Delta r_i = r_{i+1} - r_i$. (The radial separation $r$ is in units of $h^{-1}$ Mpc.)} \\
\toprule
$r_i$ & $r_{i+1}$ & fiducial & $\frac{\partial}{\partial \ln n_{\text{gal}}}$ & $\frac{\partial}{\partial \ln \sigma_{\log M}}$ & $\frac{\partial}{\partial \ln M_1 / M_{\text{min}}}$ & $\frac{\partial}{\partial \ln \alpha}$ & $\frac{\partial}{\partial Q_{\text{env}}}$ & $\frac{\partial}{\partial \Delta \gamma}$ & $\frac{\partial}{\partial \ln \Omega_M}$ & $\frac{\partial}{\partial \ln \sigma_8}$ \\
\midrule
\endfirsthead
\toprule
$r_i$ & $r_{i+1}$ & fiducial & $\frac{\partial}{\partial \ln n_{\text{gal}}}$ & $\frac{\partial}{\partial \ln \sigma_{\log M}}$ & $\frac{\partial}{\partial \ln M_1 / M_{\text{min}}}$ & $\frac{\partial}{\partial \ln \alpha}$ & $\frac{\partial}{\partial Q_{\text{env}}}$ & $\frac{\partial}{\partial \Delta \gamma}$ & $\frac{\partial}{\partial \ln \Omega_M}$ & $\frac{\partial}{\partial \ln \sigma_8}$ \\
\midrule
\endhead
0.0100 & 0.0108 & 1.2580 & 1.2753 & 1.1534 & 0.0338 & 0.5527 & -0.4272 & 2.0425 & 2.2517 & 5.5698 \\
0.0108 & 0.0117 & 1.4548 & 0.7203 & 0.7303 & 0.4045 & 1.8266 & -0.4224 & 1.0516 & -0.1883 & 3.3688 \\
0.0117 & 0.0127 & 1.0406 & 0.7779 & 0.9718 & 1.4509 & 0.8999 & 0.0453 & 1.8804 & -0.8440 & 1.1521 \\
0.0127 & 0.0137 & 1.8817 & -1.3428 & 0.7156 & 0.5894 & 0.2521 & 1.0013 & 1.8443 & 0.1055 & 1.6266 \\
0.0137 & 0.0148 & 1.4660 & -0.4338 & 0.3517 & 2.2321 & 0.8754 & -0.4500 & 1.5944 & -0.4978 & 2.5322 \\
0.0148 & 0.0160 & 1.3883 & -0.0147 & 0.1285 & 1.6188 & 0.5237 & -0.5310 & 1.5848 & 0.7348 & 1.4627 \\
0.0160 & 0.0173 & 1.4521 & -0.5605 & 1.5075 & 1.8309 & 0.5139 & 0.1016 & 1.2599 & 0.3966 & 0.6051 \\
0.0173 & 0.0188 & 1.4096 & -0.0656 & 1.2035 & 0.8554 & 0.8751 & 0.0947 & 1.0966 & -0.6393 & 2.1196 \\
0.0188 & 0.0203 & 1.6469 & -0.6154 & 0.9710 & 0.2846 & 0.6918 & 0.2325 & 2.0349 & -0.9549 & 2.4372 \\
0.0203 & 0.0219 & 1.6675 & -0.4465 & 1.0335 & 0.0435 & 0.8030 & -0.3628 & 0.8576 & 0.6676 & 1.3278 \\
0.0219 & 0.0237 & 1.6146 & -0.0417 & 0.9494 & 1.4216 & 0.9363 & -0.0042 & 1.5050 & -0.5882 & 1.2912 \\
0.0237 & 0.0257 & 1.5671 & -0.2953 & 0.8347 & 1.4575 & 0.9994 & -0.2584 & 1.5577 & 0.1676 & 0.8287 \\
0.0257 & 0.0278 & 1.7107 & -0.4526 & 0.8983 & 1.4228 & 0.6602 & -0.5380 & 1.4557 & 0.6673 & -0.2531 \\
0.0278 & 0.0301 & 1.6444 & 0.3447 & 0.9715 & 0.9072 & 0.9190 & 0.0676 & 1.1074 & 0.1264 & -0.0362 \\
0.0301 & 0.0325 & 1.5369 & 0.0745 & 0.9383 & 1.2343 & 0.7846 & 0.0871 & 1.2212 & 0.2263 & 0.2929 \\
0.0325 & 0.0352 & 1.4292 & -0.0436 & 0.8615 & 1.4501 & 0.4418 & -0.1176 & 1.1547 & 0.7006 & 1.1402 \\
0.0352 & 0.0381 & 1.5076 & -0.1815 & 0.8393 & 0.8098 & 0.9122 & 0.1970 & 1.4245 & 0.0489 & -0.1854 \\
0.0381 & 0.0412 & 1.5551 & -0.7212 & 0.7777 & 0.7426 & 0.6608 & -0.2883 & 1.0808 & -0.0123 & -0.4484 \\
0.0412 & 0.0445 & 1.4231 & -0.2364 & 0.8086 & 1.6098 & 0.6976 & -0.0522 & 1.0795 & -0.7532 & 0.2893 \\
0.0445 & 0.0482 & 1.4845 & -0.3079 & 0.9368 & 0.7263 & 0.9322 & 0.3293 & 1.1406 & 0.3317 & 0.3368 \\
0.0482 & 0.0521 & 1.4294 & -0.3724 & 0.8237 & 0.9453 & 0.8085 & 0.2867 & 0.9342 & -0.2097 & -1.2119 \\
0.0521 & 0.0564 & 1.4917 & -0.3355 & 0.9477 & 1.3287 & 0.8780 & -0.3415 & 1.1154 & 0.5451 & 0.2133 \\
0.0564 & 0.0610 & 1.3935 & -0.1602 & 0.7999 & 0.3756 & 0.6594 & 0.0172 & 1.0512 & 0.3075 & 0.4756 \\
0.0610 & 0.0660 & 1.4273 & -0.4504 & 0.9225 & 0.6652 & 0.5961 & -0.1787 & 0.8482 & -0.2733 & 0.3523 \\
0.0660 & 0.0714 & 1.3638 & -0.2741 & 0.7539 & 1.1898 & 0.7429 & 0.1909 & 0.9700 & 0.3391 & 0.3172 \\
0.0714 & 0.0772 & 1.2797 & -0.2646 & 0.9819 & 0.9526 & 0.6907 & -0.0696 & 0.9205 & 0.0277 & -0.0594 \\
0.0772 & 0.0835 & 1.3049 & -0.3402 & 0.8311 & 0.9409 & 0.8454 & 0.1193 & 0.9112 & 0.2935 & -0.5028 \\
0.0835 & 0.0904 & 1.2103 & -0.2425 & 0.8986 & 0.6506 & 0.7033 & -0.1112 & 0.5436 & 0.4681 & 0.1835 \\
0.0904 & 0.0977 & 1.2419 & -0.1216 & 1.0257 & 0.8900 & 0.5198 & 0.1309 & 0.5327 & 0.2523 & -0.2471 \\
0.0977 & 0.1057 & 1.1946 & -0.3020 & 0.8786 & 0.7854 & 0.4759 & -0.0525 & 0.7524 & 0.0572 & -0.1192 \\
0.1057 & 0.1144 & 1.1462 & -0.2978 & 0.9023 & 0.7332 & 0.5052 & -0.0451 & 0.6927 & -0.1039 & -0.3677 \\
0.1144 & 0.1237 & 1.0882 & -0.2480 & 0.7989 & 0.6466 & 0.7066 & 0.0009 & 0.5836 & 0.2169 & 0.1051 \\
0.1237 & 0.1339 & 1.0785 & -0.3483 & 0.6506 & 0.9792 & 0.5551 & 0.1625 & 0.5805 & -0.0803 & -0.0430 \\
0.1339 & 0.1448 & 1.0585 & -0.3896 & 0.6793 & 0.4431 & 0.4730 & -0.1113 & 0.4832 & 0.0149 & -0.0489 \\
0.1448 & 0.1567 & 1.0188 & -0.2237 & 0.6966 & 1.0161 & 0.5853 & 0.0340 & 0.4849 & 0.0137 & 0.0928 \\
0.1567 & 0.1695 & 0.9987 & -0.3188 & 0.6686 & 0.9309 & 0.5822 & -0.0089 & 0.3949 & 0.1464 & -0.2239 \\
0.1695 & 0.1833 & 0.9641 & -0.3508 & 0.6952 & 0.6125 & 0.5384 & -0.0319 & 0.4936 & 0.0269 & -0.1973 \\
0.1833 & 0.1983 & 0.9300 & -0.1789 & 0.7358 & 0.6697 & 0.5811 & -0.0803 & 0.4068 & 0.0104 & -0.2897 \\
0.1983 & 0.2145 & 0.8949 & -0.1989 & 0.6331 & 0.7691 & 0.5459 & -0.0133 & 0.3429 & -0.0517 & -0.6076 \\
0.2145 & 0.2321 & 0.8843 & -0.4074 & 0.6459 & 0.9382 & 0.5413 & -0.0503 & 0.2047 & 0.0711 & -0.3081 \\
0.2321 & 0.2511 & 0.8408 & -0.2008 & 0.6929 & 0.8241 & 0.5437 & -0.0376 & 0.1689 & 0.0644 & -0.5227 \\
0.2511 & 0.2716 & 0.8426 & -0.3064 & 0.6611 & 0.8037 & 0.5202 & -0.1189 & 0.3176 & 0.0164 & -0.1857 \\
0.2716 & 0.2938 & 0.7832 & -0.2371 & 0.5952 & 0.7482 & 0.5341 & -0.0732 & 0.2242 & 0.0525 & -0.5188 \\
0.2938 & 0.3178 & 0.7560 & -0.3155 & 0.5699 & 0.7351 & 0.4940 & -0.0808 & 0.1086 & 0.0919 & -0.3564 \\
0.3178 & 0.3438 & 0.7540 & -0.2818 & 0.5452 & 0.8354 & 0.4224 & -0.0250 & 0.1497 & 0.1700 & -0.4362 \\
0.3438 & 0.3719 & 0.6965 & -0.3639 & 0.5539 & 0.4298 & 0.4601 & 0.0675 & 0.1265 & 0.0562 & -0.4112 \\
0.3719 & 0.4024 & 0.6829 & -0.2194 & 0.5420 & 0.8149 & 0.4213 & -0.1475 & -0.0100 & 0.1798 & -0.3281 \\
0.4024 & 0.4353 & 0.6503 & -0.3221 & 0.4819 & 0.6676 & 0.3450 & -0.0274 & 0.0477 & 0.0308 & -0.3291 \\
0.4353 & 0.4709 & 0.5937 & -0.3700 & 0.4120 & 0.5448 & 0.4072 & -0.0543 & -0.0395 & 0.2021 & -0.3466 \\
0.4709 & 0.5094 & 0.5782 & -0.2920 & 0.4085 & 0.6102 & 0.3346 & -0.0229 & -0.0202 & 0.2432 & -0.4345 \\
0.5094 & 0.5510 & 0.5286 & -0.3660 & 0.2948 & 0.4006 & 0.3250 & 0.0219 & -0.1394 & 0.1688 & -0.3974 \\
0.5510 & 0.5961 & 0.4954 & -0.3846 & 0.2872 & 0.5136 & 0.2900 & 0.0216 & -0.1399 & 0.1629 & -0.2438 \\
0.5961 & 0.6449 & 0.4512 & -0.2883 & 0.2309 & 0.4235 & 0.2456 & -0.0546 & -0.1093 & 0.2511 & -0.1118 \\
0.6449 & 0.6976 & 0.4094 & -0.2580 & 0.2131 & 0.4108 & 0.2162 & 0.0256 & -0.0824 & 0.1667 & 0.0200 \\
0.6976 & 0.7547 & 0.3253 & -0.2716 & 0.1339 & 0.2608 & 0.2145 & 0.0404 & -0.1315 & 0.1183 & -0.0018 \\
0.7547 & 0.8164 & 0.2877 & -0.2168 & 0.0919 & 0.2653 & 0.1168 & 0.0468 & -0.1005 & 0.0782 & 0.1027 \\
0.8164 & 0.8831 & 0.2354 & -0.1778 & 0.0414 & 0.2228 & 0.0948 & 0.0913 & -0.1918 & 0.1682 & 0.0169 \\
0.8831 & 0.9554 & 0.1976 & -0.1392 & 0.0097 & 0.1452 & 0.0837 & 0.0732 & -0.1159 & 0.1599 & 0.1032 \\
0.9554 & 1.0335 & 0.1792 & -0.1339 & 0.0088 & 0.0716 & 0.0276 & 0.0916 & -0.1125 & 0.1272 & -0.0311 \\
1.0335 & 1.1180 & 0.1356 & -0.0815 & -0.0390 & 0.0743 & -0.0138 & 0.1021 & -0.1092 & 0.1061 & 0.0822 \\
1.1180 & 1.2095 & 0.1256 & -0.0754 & -0.0576 & 0.0926 & -0.0014 & 0.0967 & -0.0955 & 0.0725 & -0.0315 \\
1.2095 & 1.3084 & 0.0880 & -0.0628 & -0.1034 & 0.0202 & -0.0066 & 0.1106 & -0.0648 & 0.1334 & 0.1694 \\
1.3084 & 1.4154 & 0.0738 & -0.0589 & -0.1048 & -0.0026 & -0.0514 & 0.0989 & -0.0201 & 0.1006 & 0.1664 \\
1.4154 & 1.5312 & 0.0642 & -0.0836 & -0.0726 & -0.0743 & -0.0454 & 0.1175 & -0.0544 & -0.0357 & 0.1367 \\
1.5312 & 1.6564 & 0.0504 & -0.0454 & -0.0952 & -0.0145 & -0.0005 & 0.1121 & -0.0190 & 0.0330 & 0.1210 \\
1.6564 & 1.7918 & 0.0375 & -0.0510 & -0.0653 & -0.0208 & -0.0099 & 0.1524 & 0.0076 & 0.0058 & 0.1986 \\
1.7918 & 1.9384 & 0.0350 & -0.0144 & -0.0896 & -0.0336 & -0.0032 & 0.1086 & -0.0098 & -0.0042 & 0.0968 \\
1.9384 & 2.0969 & -0.0024 & 0.0137 & -0.0466 & -0.0450 & -0.0147 & 0.0998 & 0.0082 & 0.0692 & 0.1497 \\
2.0969 & 2.2684 & 0.0047 & -0.0184 & -0.0406 & -0.0342 & 0.0182 & 0.0839 & 0.0133 & -0.0655 & 0.0901 \\
2.2684 & 2.4539 & 0.0115 & -0.0254 & -0.0406 & -0.0582 & 0.0076 & 0.1126 & -0.0010 & -0.0158 & 0.1483 \\
2.4539 & 2.6546 & 0.0005 & -0.0007 & -0.0430 & -0.0231 & 0.0260 & 0.0833 & 0.0164 & -0.0403 & -0.0684 \\
2.6546 & 2.8717 & 0.0016 & -0.0107 & -0.0268 & -0.0185 & 0.0082 & 0.0800 & -0.0013 & -0.0323 & -0.0194 \\
2.8717 & 3.1066 & -0.0026 & -0.0046 & -0.0242 & -0.0037 & 0.0247 & 0.0679 & 0.0040 & -0.0398 & 0.0622 \\
3.1066 & 3.3607 & -0.0009 & -0.0193 & -0.0267 & -0.0097 & 0.0014 & 0.0661 & 0.0032 & 0.0972 & -0.0802 \\
3.3607 & 3.6355 & 0.0048 & -0.0182 & 0.0003 & -0.0175 & 0.0154 & 0.0618 & 0.0017 & -0.0204 & 0.0196 \\
3.6355 & 3.9329 & -0.0040 & -0.0035 & -0.0037 & -0.0129 & 0.0040 & 0.0524 & -0.0015 & 0.0041 & -0.0237 \\
3.9329 & 4.2545 & -0.0011 & -0.0130 & -0.0161 & -0.0101 & 0.0026 & 0.0646 & 0.0033 & -0.0252 & 0.0653 \\
4.2545 & 4.6025 & -0.0054 & 0.0001 & -0.0033 & -0.0105 & 0.0009 & 0.0355 & 0.0003 & 0.0683 & 0.0249 \\
4.6025 & 4.9789 & 0.0025 & -0.0052 & -0.0056 & -0.0020 & -0.0004 & 0.0462 & -0.0017 & 0.0305 & -0.0272 \\
4.9789 & 5.3861 & -0.0012 & 0.0054 & 0.0006 & 0.0163 & 0.0048 & 0.0375 & -0.0002 & -0.0099 & 0.0305 \\
5.3861 & 5.8266 & 0.0031 & 0.0072 & -0.0029 & 0.0050 & 0.0023 & 0.0557 & 0.0008 & 0.0113 & 0.0033 \\
5.8266 & 6.3031 & -0.0011 & -0.0067 & -0.0002 & 0.0005 & 0.0046 & 0.0448 & 0.0025 & 0.0038 & -0.0052 \\
6.3031 & 6.8186 & -0.0070 & 0.0023 & 0.0006 & -0.0034 & -0.0021 & 0.0254 & -0.0014 & -0.0147 & -0.0678 \\
6.8186 & 7.3763 & -0.0029 & -0.0015 & 0.0002 & 0.0174 & 0.0044 & 0.0360 & 0.0008 & 0.0391 & -0.0153 \\
7.3763 & 7.9795 & -0.0008 & 0.0076 & 0.0058 & -0.0025 & -0.0032 & 0.0114 & -0.0010 & -0.0140 & 0.0227 \\
7.9795 & 8.6321 & -0.0013 & -0.0010 & -0.0031 & -0.0071 & -0.0023 & 0.0363 & 0.0019 & -0.0182 & -0.0253 \\
8.6321 & 9.3381 & 0.0032 & 0.0018 & -0.0002 & -0.0133 & 0.0047 & 0.0183 & -0.0020 & 0.0263 & -0.0005 \\
9.3381 & 10.1018 & -0.0029 & 0.0094 & 0.0031 & -0.0093 & 0.0011 & 0.0205 & -0.0001 & -0.0230 & -0.0363 \\
10.1018 & 10.9280 & 0.0026 & 0.0014 & -0.0044 & -0.0039 & 0.0039 & 0.0173 & 0.0007 & -0.0129 & 0.0335 \\
10.9280 & 11.8218 & 0.0018 & 0.0108 & -0.0088 & -0.0097 & 0.0008 & 0.0096 & -0.0006 & -0.0068 & 0.0041 \\
11.8218 & 12.7886 & 0.0035 & -0.0043 & -0.0065 & 0.0144 & 0.0106 & 0.0131 & 0.0005 & -0.0094 & -0.0429 \\
12.7886 & 13.8345 & 0.0021 & -0.0075 & -0.0013 & -0.0044 & 0.0048 & 0.0086 & 0.0002 & -0.0203 & -0.0094 \\
13.8345 & 14.9660 & 0.0036 & -0.0025 & -0.0019 & 0.0071 & 0.0007 & 0.0020 & -0.0001 & 0.0039 & -0.0281 \\
14.9660 & 16.1900 & -0.0066 & -0.0075 & -0.0039 & 0.0054 & 0.0013 & -0.0160 & 0.0001 & -0.0093 & -0.0111 \\
16.1900 & 17.5141 & 0.0006 & -0.0109 & -0.0075 & 0.0071 & 0.0002 & -0.0090 & -0.0008 & -0.0120 & -0.0266 \\
17.5141 & 18.9465 & 0.0010 & -0.0105 & -0.0099 & -0.0123 & 0.0009 & 0.0019 & 0.0003 & 0.0359 & 0.0024 \\
18.9465 & 20.4960 & -0.0028 & -0.0120 & 0.0013 & -0.0085 & 0.0043 & -0.0119 & 0.0003 & -0.0137 & 0.0028 \\
20.4960 & 22.1723 & -0.0015 & -0.0043 & -0.0021 & -0.0044 & 0.0041 & -0.0065 & -0.0000 & -0.0056 & -0.0042 \\
22.1723 & 23.9856 & 0.0035 & -0.0154 & -0.0019 & -0.0026 & -0.0040 & 0.0110 & -0.0001 & 0.0422 & 0.0573 \\
23.9856 & 25.9473 & 0.0010 & -0.0049 & -0.0005 & 0.0086 & 0.0029 & -0.0054 & -0.0004 & 0.0109 & -0.0147 \\
25.9473 & 28.0694 & -0.0028 & -0.0016 & -0.0076 & -0.0196 & 0.0100 & 0.0062 & 0.0012 & 0.0109 & -0.1068 \\
28.0694 & 30.3650 & 0.0053 & -0.0027 & -0.0048 & 0.0266 & 0.0046 & -0.0105 & -0.0004 & 0.0156 & -0.0312 \\
30.3650 & 32.8484 & 0.0010 & -0.0136 & -0.0063 & -0.0198 & -0.0063 & -0.0068 & -0.0004 & 0.0055 & 0.0222 \\
32.8484 & 35.5349 & 0.0063 & -0.0042 & -0.0104 & -0.0115 & 0.0046 & 0.0002 & 0.0005 & 0.0232 & 0.0198 \\
35.5349 & 38.4411 & 0.0030 & -0.0105 & -0.0032 & 0.0034 & -0.0066 & -0.0072 & -0.0014 & 0.0369 & 0.0338 \\
38.4411 & 41.5850 & 0.0051 & -0.0061 & 0.0052 & 0.0133 & -0.0070 & -0.0098 & -0.0002 & -0.0039 & -0.0531 \\
41.5850 & 44.9861 & 0.0102 & -0.0118 & 0.0007 & -0.0024 & -0.0074 & 0.0080 & 0.0017 & -0.0236 & -0.0258 \\
44.9861 & 48.6653 & 0.0067 & 0.0157 & 0.0004 & -0.0448 & -0.0054 & -0.0057 & 0.0001 & -0.0441 & 0.0124 \\
48.6653 & 52.6453 & 0.0000 & 0.0000 & 0.0000 & 0.0000 & 0.0000 & 0.0000 & 0.0000 & 0.0000 & 0.0000 \\
52.6453 & 56.9509 & 0.0000 & 0.0000 & 0.0000 & 0.0000 & 0.0000 & 0.0000 & 0.0000 & 0.0000 & 0.0000 \\
56.9509 & 61.6087 & 0.0000 & 0.0000 & 0.0000 & 0.0000 & 0.0000 & 0.0000 & 0.0000 & 0.0000 & 0.0000 \\
61.6087 & 66.6473 & 0.0000 & 0.0000 & 0.0000 & 0.0000 & 0.0000 & 0.0000 & 0.0000 & 0.0000 & 0.0000 \\
66.6473 & 72.0981 & 0.0000 & 0.0000 & 0.0000 & 0.0000 & 0.0000 & 0.0000 & 0.0000 & 0.0000 & 0.0000 \\
72.0981 & 77.9946 & 0.0000 & 0.0000 & 0.0000 & 0.0000 & 0.0000 & 0.0000 & 0.0000 & 0.0000 & 0.0000 \\
77.9946 & 84.3734 & 0.0000 & 0.0000 & 0.0000 & 0.0000 & 0.0000 & 0.0000 & 0.0000 & 0.0000 & 0.0000 \\
84.3734 & 91.2738 & 0.0000 & 0.0000 & 0.0000 & 0.0000 & 0.0000 & 0.0000 & 0.0000 & 0.0000 & 0.0000 \\
91.2738 & 98.7387 & 0.0000 & 0.0000 & 0.0000 & 0.0000 & 0.0000 & 0.0000 & 0.0000 & 0.0000 & 0.0000 \\
98.7387 & 106.8140 & 0.0000 & 0.0000 & 0.0000 & 0.0000 & 0.0000 & 0.0000 & 0.0000 & 0.0000 & 0.0000 \\
106.8140 & 115.5498 & 0.0000 & 0.0000 & 0.0000 & 0.0000 & 0.0000 & 0.0000 & 0.0000 & 0.0000 & 0.0000 \\
115.5498 & 125.0000 & 0.0000 & 0.0000 & 0.0000 & 0.0000 & 0.0000 & 0.0000 & 0.0000 & 0.0000 & 0.0000 \\
\bottomrule
\end{longtabu}

\begin{longtabu}{*{5}{X[1,r]}}
\caption{Logarithm of the nonlinear-to-linear matter correlation function ratio $\ln b_{\text{nl}}(r)$ and its partial derivatives with respect to cosmological parameters, tabulated in radial bins of width $\Delta r_i = r_{i+1} - r_i$. (The radial separation $r$ is in units of $h^{-1}$ Mpc.)} \\
\toprule
   $r_i$ & $r_{i+1}$ & fiducial & $\frac{\partial}{\partial \ln \sigma_8}$ & $\frac{\partial}{\partial \ln \Omega_M}$ \\
\midrule
\endfirsthead
\toprule
   $r_i$ & $r_{i+1}$ & fiducial & $\frac{\partial}{\partial \ln \sigma_8}$ & $\frac{\partial}{\partial \ln \Omega_M}$ \\
\midrule
\endhead
0.0100 & 0.0108 & 1.2705 & 0.1940 & -1.2859 \\
0.0108 & 0.0117 & 1.2750 & -1.9267 & 1.7964 \\
0.0117 & 0.0127 & 1.5851 & 0.6828 & 0.0321 \\
0.0127 & 0.0137 & 0.9493 & -1.0000 & -0.7316 \\
0.0137 & 0.0148 & 1.3020 & -1.2130 & 0.4438 \\
0.0148 & 0.0160 & 1.3643 & -0.5295 & -0.9976 \\
0.0160 & 0.0173 & 1.3980 & 0.0776 & -0.1493 \\
0.0173 & 0.0188 & 1.4122 & -1.0000 & 0.9292 \\
0.0188 & 0.0203 & 1.2505 & -1.1767 & 0.2480 \\
0.0203 & 0.0219 & 1.1800 & 0.3879 & -1.1004 \\
0.0219 & 0.0237 & 1.3508 & -0.2429 & -0.2244 \\
0.0237 & 0.0257 & 1.3452 & -0.0733 & -0.3100 \\
0.0257 & 0.0278 & 1.2719 & 0.1582 & -0.3724 \\
0.0278 & 0.0301 & 1.3697 & 0.8899 & -0.1671 \\
0.0301 & 0.0325 & 1.4595 & 0.4631 & 0.1178 \\
0.0325 & 0.0352 & 1.4280 & -0.2873 & 0.0077 \\
0.0352 & 0.0381 & 1.4149 & 0.1355 & -0.2667 \\
0.0381 & 0.0412 & 1.4253 & 0.5451 & -0.0579 \\
0.0412 & 0.0445 & 1.4691 & -0.2181 & 0.0557 \\
0.0445 & 0.0482 & 1.4490 & -0.0248 & -0.4727 \\
0.0482 & 0.0521 & 1.4735 & 0.0989 & -0.4086 \\
0.0521 & 0.0564 & 1.4405 & 0.0807 & -0.0199 \\
0.0564 & 0.0610 & 1.4561 & 0.1287 & -0.1452 \\
0.0610 & 0.0660 & 1.4240 & -0.0627 & -0.1162 \\
0.0660 & 0.0714 & 1.4458 & 0.0387 & -0.2616 \\
0.0714 & 0.0772 & 1.4418 & 0.0677 & -0.0764 \\
0.0772 & 0.0835 & 1.4350 & 0.1839 & -0.2399 \\
0.0835 & 0.0904 & 1.4350 & 0.1052 & -0.3563 \\
0.0904 & 0.0977 & 1.4316 & 0.3006 & -0.3698 \\
0.0977 & 0.1057 & 1.4397 & -0.0332 & -0.2839 \\
0.1057 & 0.1144 & 1.4409 & 0.0114 & -0.1865 \\
0.1144 & 0.1237 & 1.4179 & 0.0119 & -0.2664 \\
0.1237 & 0.1339 & 1.4092 & 0.1130 & -0.0914 \\
0.1339 & 0.1448 & 1.3983 & 0.1356 & -0.2503 \\
0.1448 & 0.1567 & 1.3810 & 0.1673 & -0.1428 \\
0.1567 & 0.1695 & 1.3626 & 0.0183 & -0.1714 \\
0.1695 & 0.1833 & 1.3502 & 0.1579 & -0.2242 \\
0.1833 & 0.1983 & 1.3297 & 0.1866 & -0.1739 \\
0.1983 & 0.2145 & 1.3111 & 0.1749 & -0.1980 \\
0.2145 & 0.2321 & 1.2917 & 0.1305 & -0.2316 \\
0.2321 & 0.2511 & 1.2717 & 0.2038 & -0.2215 \\
0.2511 & 0.2716 & 1.2466 & 0.1357 & -0.1790 \\
0.2716 & 0.2938 & 1.2282 & 0.1939 & -0.2294 \\
0.2938 & 0.3178 & 1.1998 & 0.2212 & -0.2267 \\
0.3178 & 0.3438 & 1.1765 & 0.2282 & -0.2205 \\
0.3438 & 0.3719 & 1.1463 & 0.2500 & -0.1820 \\
0.3719 & 0.4024 & 1.1162 & 0.3238 & -0.2073 \\
0.4024 & 0.4353 & 1.0878 & 0.2863 & -0.1647 \\
0.4353 & 0.4709 & 1.0557 & 0.3228 & -0.1996 \\
0.4709 & 0.5094 & 1.0236 & 0.3466 & -0.1665 \\
0.5094 & 0.5510 & 0.9907 & 0.3525 & -0.2016 \\
0.5510 & 0.5961 & 0.9590 & 0.3956 & -0.1927 \\
0.5961 & 0.6449 & 0.9218 & 0.4109 & -0.1920 \\
0.6449 & 0.6976 & 0.8831 & 0.4501 & -0.1749 \\
0.6976 & 0.7547 & 0.8422 & 0.4319 & -0.1518 \\
0.7547 & 0.8164 & 0.8014 & 0.4732 & -0.1620 \\
0.8164 & 0.8831 & 0.7598 & 0.5103 & -0.1783 \\
0.8831 & 0.9554 & 0.7146 & 0.5176 & -0.1517 \\
0.9554 & 1.0335 & 0.6664 & 0.5795 & -0.1437 \\
1.0335 & 1.1180 & 0.6211 & 0.5612 & -0.1388 \\
1.1180 & 1.2095 & 0.5696 & 0.6113 & -0.1469 \\
1.2095 & 1.3084 & 0.5241 & 0.6069 & -0.1330 \\
1.3084 & 1.4154 & 0.4727 & 0.6359 & -0.1307 \\
1.4154 & 1.5312 & 0.4235 & 0.6381 & -0.1410 \\
1.5312 & 1.6564 & 0.3721 & 0.6082 & -0.1521 \\
1.6564 & 1.7918 & 0.3228 & 0.5731 & -0.1401 \\
1.7918 & 1.9384 & 0.2764 & 0.5147 & -0.1351 \\
1.9384 & 2.0969 & 0.2326 & 0.4594 & -0.1329 \\
2.0969 & 2.2684 & 0.1933 & 0.4187 & -0.1295 \\
2.2684 & 2.4539 & 0.1556 & 0.3264 & -0.1311 \\
2.4539 & 2.6546 & 0.1222 & 0.2765 & -0.1168 \\
2.6546 & 2.8717 & 0.0962 & 0.2051 & -0.1260 \\
2.8717 & 3.1066 & 0.0705 & 0.1560 & -0.1182 \\
3.1066 & 3.3607 & 0.0520 & 0.0799 & -0.0966 \\
3.3607 & 3.6355 & 0.0340 & 0.0478 & -0.0825 \\
3.6355 & 3.9329 & 0.0196 & 0.0371 & -0.0930 \\
3.9329 & 4.2545 & 0.0065 & 0.0079 & -0.0711 \\
4.2545 & 4.6025 & -0.0004 & 0.0033 & -0.0637 \\
4.6025 & 4.9789 & -0.0105 & -0.0099 & -0.0611 \\
4.9789 & 5.3861 & -0.0160 & -0.0245 & -0.0620 \\
5.3861 & 5.8266 & -0.0221 & -0.0800 & -0.0644 \\
5.8266 & 6.3031 & -0.0296 & -0.0453 & -0.0546 \\
6.3031 & 6.8186 & -0.0345 & -0.0690 & -0.0429 \\
6.8186 & 7.3763 & -0.0364 & -0.0713 & -0.0316 \\
7.3763 & 7.9795 & -0.0397 & -0.0645 & -0.0318 \\
7.9795 & 8.6321 & -0.0437 & -0.1004 & -0.0247 \\
8.6321 & 9.3381 & -0.0462 & -0.0967 & -0.0247 \\
9.3381 & 10.1018 & -0.0470 & -0.0918 & -0.0317 \\
10.1018 & 10.9280 & -0.0456 & -0.0918 & -0.0305 \\
10.9280 & 11.8218 & -0.0472 & -0.0947 & -0.0056 \\
11.8218 & 12.7886 & -0.0485 & -0.0563 & -0.0146 \\
12.7886 & 13.8345 & -0.0455 & -0.0326 & 0.0028 \\
13.8345 & 14.9660 & -0.0433 & -0.0387 & -0.0106 \\
14.9660 & 16.1900 & -0.0388 & -0.0465 & -0.0186 \\
16.1900 & 17.5141 & -0.0336 & -0.0326 & 0.0025 \\
17.5141 & 18.9465 & -0.0300 & -0.0680 & -0.0028 \\
18.9465 & 20.4960 & -0.0289 & -0.0635 & -0.0077 \\
20.4960 & 22.1723 & -0.0221 & -0.0254 & 0.0305 \\
22.1723 & 23.9856 & -0.0243 & -0.0991 & 0.0492 \\
23.9856 & 25.9473 & -0.0240 & -0.0875 & 0.0139 \\
25.9473 & 28.0694 & -0.0271 & -0.1361 & 0.0443 \\
28.0694 & 30.3650 & -0.0240 & -0.0804 & 0.0100 \\
30.3650 & 32.8484 & -0.0238 & -0.0449 & 0.0143 \\
32.8484 & 35.5349 & -0.0055 & 0.0252 & 0.0041 \\
35.5349 & 38.4411 & 0.0049 & -0.0603 & -0.0126 \\
38.4411 & 41.5850 & 0.0119 & -0.0598 & 0.0128 \\
41.5850 & 44.9861 & 0.0183 & 0.0050 & -0.0088 \\
44.9861 & 48.6653 & 0.0152 & -0.1258 & -0.1206 \\
48.6653 & 52.6453 & -0.0013 & -0.2473 & -0.0331 \\
52.6453 & 56.9509 & 0.0267 & -0.2889 & -0.0506 \\
56.9509 & 61.6087 & 0.0561 & -0.2965 & -0.0521 \\
61.6087 & 66.6473 & 0.1409 & -0.2396 & -0.2008 \\
66.6473 & 72.0981 & 0.2364 & -0.3476 & 0.0904 \\
72.0981 & 77.9946 & 0.2831 & 0.0611 & 0.3983 \\
77.9946 & 84.3734 & 0.3476 & -0.7497 & 0.4894 \\
84.3734 & 91.2738 & 0.4124 & -0.1305 & -1.3700 \\
91.2738 & 98.7387 & 0.1004 & 0.1881 & -1.6015 \\
98.7387 & 106.8140 & -0.0989 & -0.1172 & 0.1546 \\
106.8140 & 115.5498 & -0.1862 & 0.1121 & -0.2808 \\
115.5498 & 125.0000 & 0.3549 & 2.7004 & 0.2328 \\
\bottomrule
\end{longtabu}


\bsp	
\label{lastpage}
\end{document}